\documentclass[aps,prl,twocolumn,superscriptaddress,showpacs]{revtex4-1}
\usepackage{amsmath,amssymb,graphics,epsfig,epstopdf,color,verbatim,ulem,braket,tabularx}
\usepackage[colorlinks,linkcolor=blue,citecolor=blue,urlcolor=blue]{hyperref}

\graphicspath{{Figures/}}

\newcommand\COMMENTED[1] {}

\begin{document}

\title{Precision Many-Body Study of the Berezinskii-Kosterlitz-Thouless Transition and Temperature-Dependent Properties in the Two-Dimensional Fermi Gas}

\author{Yuan-Yao He}
\email{heyuanyao@nwu.edu.cn}
\affiliation{Institute of Modern Physics, Northwest University, Xi'an 710127, China}
\affiliation{Shaanxi Key Laboratory for Theoretical Physics Frontiers, Xi'an 710127, China}
\affiliation{Center for Computational Quantum Physics, Flatiron Institute, New York, New York 10010, USA}

\author{Hao Shi}
\affiliation{Department of Physics and Astronomy, University of Delaware, Newark, Delaware 19716, USA}

\author{Shiwei Zhang}
\email{szhang@flatironinstitute.org}
\affiliation{Center for Computational Quantum Physics, Flatiron Institute, New York, New York 10010, USA}

\begin{abstract}
We perform large-scale, numerically exact calculations on the two-dimensional interacting Fermi gas with a contact attraction. Reaching much larger lattice sizes and lower temperatures than previously possible, we determine systematically the finite-temperature phase diagram of the Berezinskii-Kosterlitz-Thouless (BKT) transitions for interaction strengths ranging from BCS to crossover to BEC regimes. The evolutions of the pairing wave functions and the fermion and Cooper pair momentum distributions with temperature are accurately characterized. In the crossover regime, we find that the contact has a nonmonotonic temperature dependence, first increasing as temperature is lowered, and then showing a slight decline below the BKT transition temperature to approach the ground-state value from above. 
\end{abstract}


\date{\today}
\maketitle

Two-dimensional (2D) correlated fermion systems have been of central interest in condensed matter physics and other areas. They vary from lattice models~\cite{LeBlanc2015,Arovas2021} to ultracold quantum gas~\cite{Bloch2008,*Giorgini2008} and real materials~\cite{Castro2009,*Kotov2012}. The interplay between the reduced dimensionality and many-body correlation effects in such systems can induce fascinating and unique quantum phenomena, such as Berezinskii-Kosterlitz-Thouless (BKT) phase transitions~\cite{Berezinsky1972,Kosterlitz1973,Jorge2013,Ryzhov2017} and  high-temperature superconductivity~\cite{Patrick2006}. Among them, the 2D Fermi gas with zero-range attractive interaction presents tremendous opportunities as it can be experimentally realized using ultracold atoms~\cite{Bloch2008,*Giorgini2008} in a highly controlled way. The system has already contributed greatly to our understanding of BCS-BEC crossover physics~\cite{Eagles1969,Schmitt1985,Randeria1989,Randeria2014,Giancarlo2018}. With intense ongoing effort and rapid experimental 
advances, it is poised to play an even greater role in the quest to understand the physics of 2D correlated fermion systems.

In experiments, the interaction strength of the 2D Fermi gas can be tuned by the scattering length via Feshbach resonance, and a wide range of fermion densities and temperatures can be accessed with great control and precision. Such a dilute Fermi gas system was first realized with a harmonic trap~\cite{Martiyanov2010,Feld2011,Alexey2011,Dyke2011,Sommer2012,Zhang2012,Koschorreck2012,Vogt2012,Giamarchi2012,Makhalov2014,Ries2015,Murthy2015,Dyke2016,Boettcher2016,Mitra2016,Fenech2016,Holten2018,Murthy2018,Peppler2018} and recently in a box potential~\cite{Hueck2018,Sobirey2021}. Early experiments studied the density distribution in a trap~\cite{Alexey2011}, the 2D-3D crossover~\cite{Dyke2011,Sommer2012}, polarons~\cite{Zhang2012,Koschorreck2012}, viscosity~\cite{Vogt2012}, the contact parameter~\cite{Giamarchi2012}, pressure~\cite{Makhalov2014}, followed by the equation of state~\cite{Boettcher2016,Fenech2016}. Superfluidity at low temperatures and the corresponding BKT phase transitions were observed through measurements of the pair condensate~\cite{Ries2015}, first-order correlation function $g_1(r)$~\cite{Murthy2015} and the critical velocity~\cite{Sobirey2021}, and the BKT transition temperature in the crossover regime was measured in Ref.~\onlinecite{Ries2015}. Many more experiments can be expected, with increasing capability, precision, and control. This has stimulated much theoretical and computational activity and opened an avenue for rapid progress through comparison and benchmark. 

Theoretically, the 2D Fermi gas is usually described by a model including a simple dispersion (e.g., quadratic) and contact attraction~\cite{Turlapov2017}. A variety of approximate theories have been applied, including mean-field analysis~\cite{Miyake1983,*Schmitt1989,*Randeria1989,*Randeria1990,*Drechsler1992,*Petrov2003,*Salasnich2007,*Zhang2008,*Lianyi2015,*Bighin2016}, virial expansion~\cite{Drummond2010,Jesper2013}, and the Luttinger-Ward approach~\cite{Watanabe2013,Bauer2014,Mulkerin2015,Matsumoto2017}. These studies have concentrated on the equation of state, pair correlations, BKT transitions and the possible pseudogap phenomena. Computationally, ground-state properties have been characterized reasonably well, with fixed-node diffusion Monte Carlo simulations\cite{Bertaina2011,Alexandros2016} and numerically exact auxiliary-field quantum Monte Carlo (AFQMC) method~\cite{Shihao2015,Vitali2017}. At finite temperatures, a quantum Monte Carlo (QMC) study employing state-of-the-art lattice techniques provided numerically exact results on the pressure, compressibility and the contact~\cite{Anderson2015}; however these simulations were still mostly limited to finite lattices of $\sim 400$ sites and in the normal phase at higher temperatures.

In this Letter, we report an {\it ab initio}, numerically exact study of the finite-temperature properties of the BCS-BEC crossover in the 2D Fermi gas. Implementing recent progress in AFQMC~\cite{YuanYao2019}, our calculations reach lattice sizes ($\sim 5000$ sites) and temperatures ($T_F\sim 0.0125\,T_F$) far beyond what has been possible with existing methods. This allows us to approach the continuum and thermodynamic limits, and compute quantities previously inaccessible from simulations or determine properties with much higher precision. We obtain the phase diagram of the BKT transition, and characterize the evolution of the pairing wave functions and the fermion and pair momentum distributions. An accurate measure of the contact is provided. 

We model the uniform 2D Fermi gas with contact attraction by the following lattice Hamiltonian,
\begin{equation}
\label{eq:HubbardModel}
\begin{split}
\hat{H} = 
\sum_{\mathbf{k}\sigma}\varepsilon_{\mathbf{k}}c_{\mathbf{k}\sigma}^+c_{\mathbf{k}\sigma}
+ U\sum_{\mathbf{i}}\hat{n}_{\mathbf{i}\uparrow}\hat{n}_{\mathbf{i}\downarrow}
- \mu\sum_{\mathbf{i},\sigma} \hat{n}_{\mathbf{i}\sigma},
\end{split}
\end{equation}
where $\sigma$ ($=\uparrow$ or $\downarrow$) denotes spin, and $\hat{n}_{\mathbf{i}\sigma}=c_{\mathbf{i}\sigma}^+c_{\mathbf{i}\sigma}$ is the density operator. 
We have tested both the Hubbard $\varepsilon_{\mathbf{k}}=4-2(\cos k_x+\cos k_y)$ and the quadratic dispersions $\varepsilon_{\mathbf{k}}=k_x^2+k_y^2$ (corresponding to fermion mass $m=1/2$ comparing to $\varepsilon_{\mathbf{k}}=\hbar^2k^2/2m$), where the momentum $k_x$ (and $k_y$) are defined in units of $2\pi/L$ with the system size $N_s=L^2$. These dispersions, which both have finite effective ranges~\cite{Carlson2011} that vanish as $L\to\infty$, lead to consistent results in the large $L$ limit. The Hubbard dispersion tends to have larger finite-size effects, which are more prominent in the contact. We use it for cross-checks, but all our final results are obtained with the quadratic dispersion. In practical simulations, the chemical potential $\mu$ is tuned to reach a fixed number of fermions $N_e$, resulting in a fermion density $n=N_e/N_s$. The on-site interaction strength $U$ can be determined from $\log(k_Fa)$~\cite{Werner2012,Shihao2015,YuanYao2019} (with the Fermi vector $k_F=\sqrt{2\pi n}$ and the 2D scattering length $a$). We measure temperatures in units of $T_F\equiv E_F/k_B$ (setting $k_B=1$) with the Fermi energy $E_F=k_F^2=2\pi n$. To reach the continuum limit reliably, especially given the delicate nature of the BKT transition, we have simulated lattice sizes  up to $75\times 75$. To span the temperature range and make connection with the ground state, we access temperatures as low as $T/T_F=0.0125$.

Perhaps the most intriguing property of 2D Fermi gas is the BKT phase transition~\cite{Berezinsky1972,Kosterlitz1973}. We compute the transition temperatures $T_{\text{BKT}}/T_F$ numerically from the condensate fraction in finite systems. The condensate fraction, $n_c$, is obtained as the leading eigenvalue of the zero-momentum spin-singlet pairing matrix~\cite{Shihao2015,YuanYao2019}
\begin{equation}
\label{eq:Qeq0PairingMat}
\begin{split}
\mathbf{M}_{\mathbf{k}\mathbf{k}^{\prime}}
=\langle\Delta_{\mathbf{k}}^+\Delta_{\mathbf{k}^{\prime}}\rangle-\delta_{\mathbf{k}\mathbf{k}^{\prime}}\langle c_{\mathbf{k}\uparrow}^+c_{\mathbf{k}\uparrow}\rangle\langle c_{-\mathbf{k}\downarrow}^+c_{-\mathbf{k \downarrow}}\rangle,
\end{split}
\end{equation}
divided by $N_e/2$, with $\Delta_{\mathbf{k}}^+=c_{\mathbf{k}\uparrow}^+c_{-\mathbf{k}\downarrow}^+$ as the pairing operator. The corresponding eigenvector gives the pairing wave function in reciprocal space $\phi_{\uparrow\downarrow}(\mathbf{k})$. In a finite-size system the first-order derivative $d\,n_c/d\,(T/T_F)$ shows a sharp peak~\cite{YuanYao2019}, whose location identifies the BKT transition. We fit $n_c(T)$ in each system using a fourth-order polynomial around the transition point, and then compute the peak location of its first-order derivative. We then perform a finite-size  extrapolation to obtain $T_{\text{BKT}}$ in the thermodynamic limit~\cite{Suppl}. (We have also tested using the first-order correlation function  and studying its decay exponent, and find the finite-size effects are much larger~\cite{Moreo1991,Santos1993,Paiva2004,Nakano2006,Shailesh2002}.) Systematic errors from finite-size effects are removed or estimated from the extrapolation process~\cite{Suppl}. Other systematic errors (Trotter errors, truncation errors) are controlled and smaller than our statistical uncertainties. The latter are estimated from the Monte Carlo process as one standard deviation errors.

\begin{figure}[t]
\centering
\includegraphics[width=0.98\columnwidth]{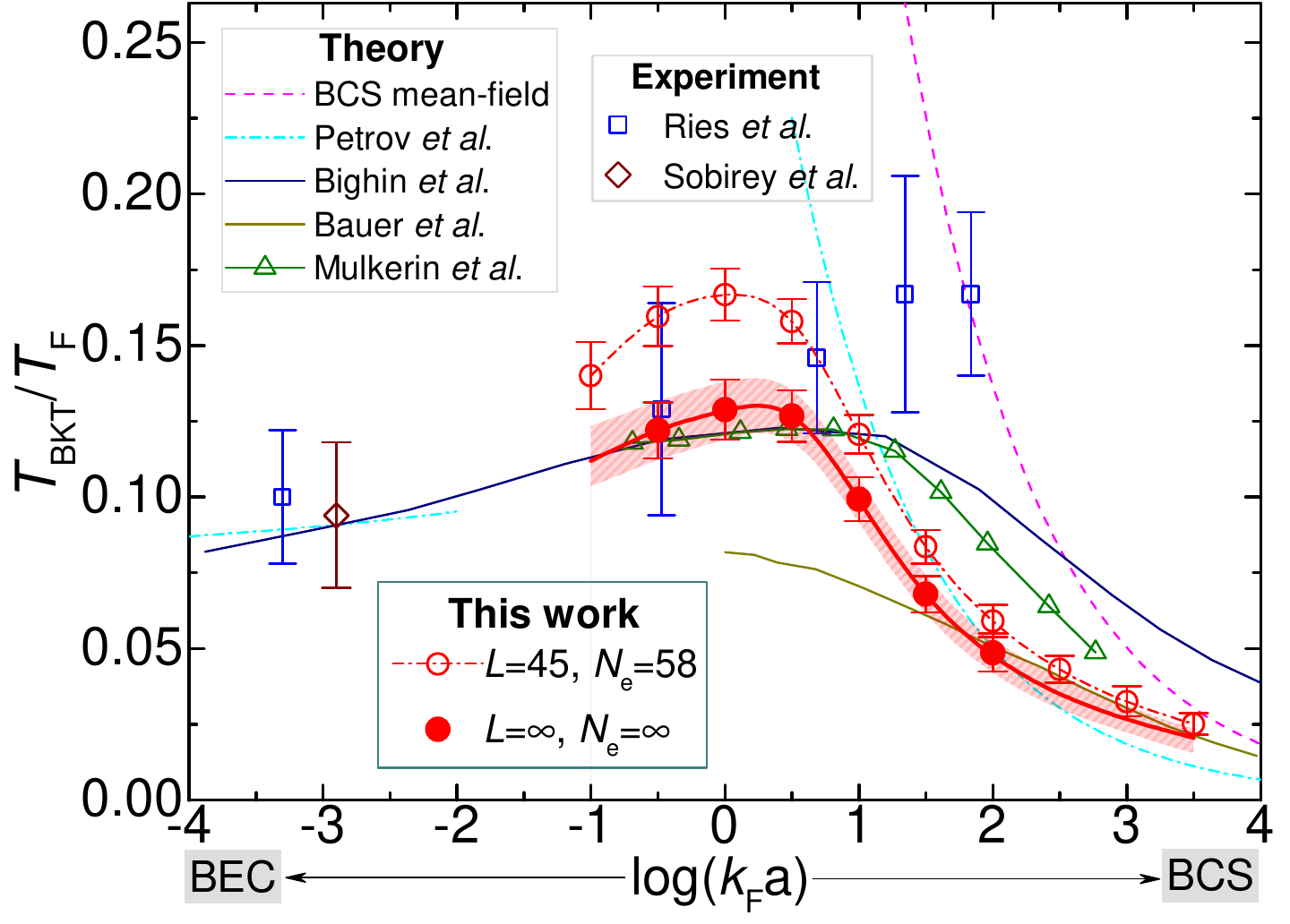}
\caption{\label{fig:BKTPhaseDiagram} BKT transition temperatures and phase diagram of the 2D interacting Fermi gas. Empty red circles show our exact results for a finite system of $L=45, N_e=58$ with quadratic dispersion. Filled red circles show finite-size scaling results to the continuum and thermodynamic limits ($L=\infty, N_e=\infty$) for a subset of the interaction strengths. The solid red line connecting these are the results of interpolation, with the shaded band indicating statistical error bars based on both sets of results. For comparison, results are also shown from BCS mean-field theory and its improvement on the BCS side (Petrov {\it et~al.}~\cite{Petrov2003}), the weakly interacting Bose gas on the BEC side (Petrov {\it et~al.}~\cite{Petrov2003}), one-loop Gaussian fluctuation theory (Bighin {\it et~al.}~\cite{Bighin2016}), Luttinger-Ward theory (Bauer {\it et~al.}~\cite{Bauer2014}), Gaussian pair fluctuation theory (Mulkerin {\it et~al.}~\cite{Mulkerin2017}), and experimental measurements (Ries {\it et~al.}~\cite{Ries2015} and Sobirey {\it et~al.}~\cite{Sobirey2021}).  }
\end{figure}

\begin{figure*}[t!]
\centering
\includegraphics[width=0.98\textwidth]{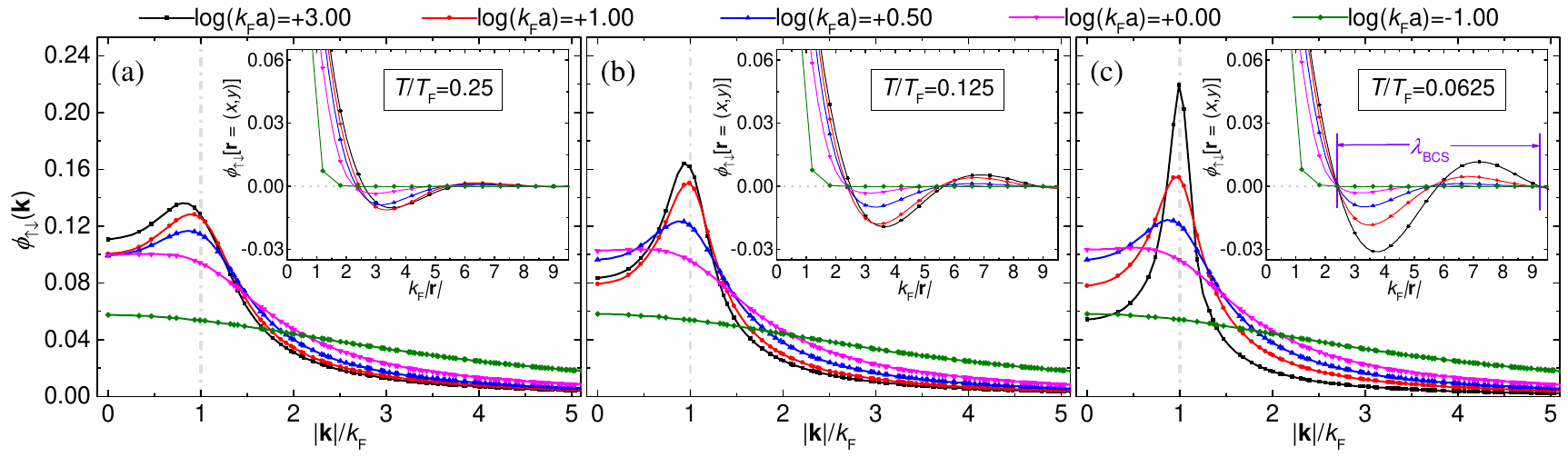}
\caption{\label{fig:RKPairWvfc} The singlet pairing wave function in reciprocal space versus $|\mathbf{k}|/k_F$ (main plots), and its Fourier transform, the real space pairing wave function plotted versus $k_F|\mathbf{r}|$ along the diagonal line ($x=y$) (insets). Three temperatures are displayed, from high to low, in the three panels from left to right: (a) $T/T_F=0.25$, (b) $T/T_F=0.125$, and (c) $T/T_F=0.0625$. In each panel five $\log(k_Fa)$ values are shown, as indicated by the legends on top, to span the entire range of the interaction strength. The statistical uncertainties are smaller than the symbol size in all cases and are not shown. In the main plots, the Fermi surface is indicated by a vertical dot-dashed line. These calculations are performed with $L=45$ and $N_e=58$.}
\end{figure*}

Our main results are summarized in Fig.~\ref{fig:BKTPhaseDiagram}, which presents the phase diagram of the 2D Fermi gas. The BKT transition temperatures $T_{\text{BKT}}/T_F$ are obtained from an extensive set of individual AFQMC calculations, yielding numerically exact solutions for the Hamiltonian $\hat H(N_s,N_e)$ on finite lattices. As mentioned, the most accurate finite-temperature results from previous QMC calculations~\cite{Anderson2015} were limited to high temperatures, mostly  in the normal phase. Our AFQMC calculations, employing several methodological advances including a low-rank factorization technique, are able to study both normal and superfluid phases by reaching much lower temperatures, and compute properties accurately at the continuum limit by reaching much larger lattice sizes ($N_s\sim 5000$ vs.~$N_s<400$). This allows reliable finite-size extrapolation to $N_s\rightarrow \infty$~\cite{Tomita2002,Filinov2010,Nguyen2019}. Results of $T_{\text{BKT}}/T_F$ are shown for a gas of $L=45, N_e=58$ for ten interaction strengths spanning the BCS-BEC crossover. Then for selected interaction strengths, we perform systematic finite-size scaling with a range of $N_e$ values (up to $122$)~\cite{Suppl},  each at the continuum limit, to estimate $T_{\text{BKT}}/T_F$ at the thermodynamic limit $N_e=\infty$.  

The highest BKT transition temperature occurs in the crossover regime, for example  at $\log(k_Fa)=+0.0$, $T_{\rm BKT}/T_F=0.129(9)$ (statistically indistinguishable from the predicted upper bound of $0.125$~\cite{Hazra2019}). This is likely the balance between competing trends. In the weak coupling BCS regime, $T_{\text{BKT}}/T_F$ decreases with interaction. On the other hand, as interaction is increased, it is observed that the Cooper pairs become more massive~\cite{Giamarchi2012}, suppressing the tendency for phase coherence. The $T_{\text{BKT}}/T_F$ values measured from experiment~\cite{Ries2015} are also shown in Fig.~\ref{fig:BKTPhaseDiagram}. Experimental error bars are still large but our results are consistent with the measured results in the BEC and crossover regimes. In a more recent experiment~\cite{Sobirey2021} the BKT transition temperature is measured in the deep BEC regime [$\log(k_Fa)=-2.9$] to be $0.094(24)$. Although this is well outside the interaction strength we studied, the result seems compatible with the trend in our curve at the smallest $\log(k_Fa)$ value. A clear discrepancy is seen in the BCS regime between our results and experiment. This could be due to the experimental analysis procedure~\cite{Matsumoto2016}. Further investigations are needed which will undoubtedly lead to major progress in this important problem.

In Fig.~\ref{fig:RKPairWvfc},  we show the evolution of the spin-singlet pairing wave function at three temperatures. At low temperature, $T/T_F=0.0625$ [panel (c)], the system is inside or close to the superfluid phase, and the results are quantitatively close to the ground state values~\cite{Shihao2015}, which provide a consistency check. In the BCS regime [$\log(k_Fa)=+3.00$], the pairing wave function shows a sharp peak around the Fermi surface, and in real space it extends through the whole system with a wave of approximate wavelength $\lambda_{BCS}=2\pi/k_F$, indicating a Cooper pair with size comparable to the system size. In the BEC regime [$\log(k_Fa)=-1.00$], the pairing wave function becomes rather flat  in reciprocal space, and tightly bound in real space with size much smaller than the interparticle spacing $1/k_F$. Between these limits, the pairing wave function provides a visualization of the crossover process. As the interaction strength increases, the wave function smoothly evolves across a strongly interacting ``unitary'' regime, with the peak at short distance growing rapidly and the tail of the real-space wave function decaying correspondingly. 

Towards higher temperatures $T/T_F=0.125$ and $T/T_F=0.25$ as shown in Figs.~\ref{fig:RKPairWvfc}(b) and ~\ref{fig:RKPairWvfc}(a), the peak of the pairing wave function $\phi_{\uparrow\downarrow}(\mathbf{k})$ is suppressed. Correspondingly the tail of the real-space wave function is seen to decay significantly as $T/T_F$ is increased.This trend is more pronounced in the BCS regime at $\log(k_Fa)=+3.00$ (black lines). The evolution of the pairing wave function versus  temperatures in the crossover and BEC regimes are more gradual. The difference in the pairing wave function appears to be quite mild with respect to whether the system is in the normal or superfluid phase, i.e., whether above or below the transition temperature given in Fig.~\ref{fig:BKTPhaseDiagram}. (A larger difference is seen in the finite-size condensate fraction~\cite{Suppl}.)

\begin{figure*}
\centering
\includegraphics[width=0.98\textwidth]{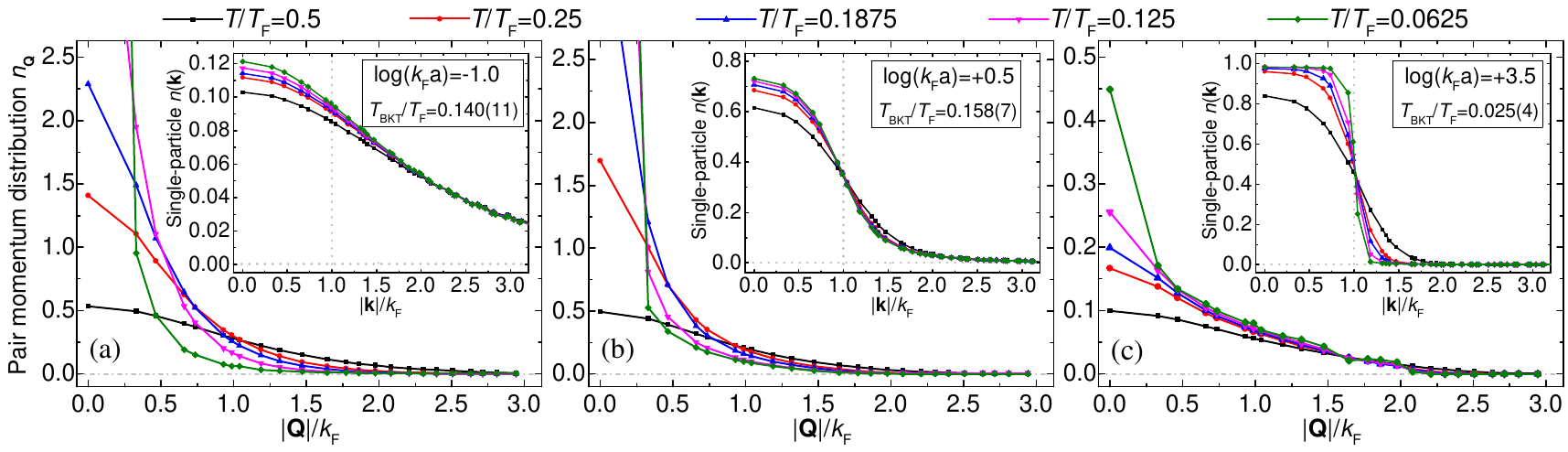}
\caption{\label{fig:MomentDistrit} The momentum distributions for Cooper pairs $n_{\mathbf{Q}}$ versus $|\mathbf{Q}|/k_F$ (main plot) and fermions $n(\mathbf{k})$ versus $|\mathbf{k}|/k_F$ (insets), for three interaction strengths: (a) $\log(k_Fa)=-1.0$ in BEC regime, (b) $\log(k_Fa)=+0.5$ in crossover regime, and (c) $\log(k_Fa)=+3.5$ in BCS regime, with temperatures $T/T_F=0.0625 \sim 0.5$. These calculations are performed with $L=45$ and $N_e=58$, and the corresponding $T_{\rm BKT}/T_F$ values from Fig.~\ref{fig:BKTPhaseDiagram} are indicated.}
\end{figure*}

The momentum distributions for fermions and Cooper pairs can be measured in experiments~\cite{Ries2015}. To allow direct comparisons, we have computed both quantities in our AFQMC calculations. The pair momentum distribution is obtained by extending the pairing matrix in Eq.~(\ref{eq:Qeq0PairingMat}) to Cooper pairs with finite center-of-mass momentum $\mathbf{Q}$ as
\begin{equation}
\label{eq:QPairingMat}
\begin{split}
\mathbf{M}_{\mathbf{k}\mathbf{k}^{\prime},\mathbf{Q}}
=\langle\Delta_{\mathbf{k}\mathbf{Q}}^+\Delta_{\mathbf{k}^{\prime}\mathbf{Q}}\rangle-\delta_{\mathbf{k}\mathbf{k}^{\prime}}\langle c_{\mathbf{k}+\mathbf{Q}\uparrow}^+c_{\mathbf{k}+\mathbf{Q}\uparrow}\rangle\langle c_{-\mathbf{k}\downarrow}^+c_{-\mathbf{k \downarrow}}\rangle,
\end{split}
\end{equation}
with $\Delta_{\mathbf{k}\mathbf{Q}}^+=c_{\mathbf{k}+\mathbf{Q}\uparrow}^+c_{-\mathbf{k}\downarrow}^+$. At each $\mathbf{Q}$, we measure the $\mathbf{M}_{\mathbf{k}\mathbf{k}^{\prime},\mathbf{Q}}$ matrix, and its leading eigenvalue is identified as the pair momentum distribution $n_{\mathbf{Q}}$. Thus, $n_{\mathbf{Q}=0}$ recovers the condensate fraction result discussed earlier. The results for $n_{\mathbf{Q}}$ are shown in Fig.~\ref{fig:MomentDistrit}, for three representative interactions. In each case, the pair momentum distribution becomes rapidly centered at $\mathbf{Q}=0$ as the temperature is decreased. This behavior is consistent with that of a system of interacting bosonic Cooper pairs, in which only the $\mathbf{Q}=0$ component  will survive in the ground state in the bulk limit. At finite temperatures, some of the Cooper pairs can either be broken into individual fermions or simply acquire a velocity (momentum), turning into finite center-of-mass momentum pairs~\cite{Qijin1998}. Furthermore, we find that $\ln n_{\mathbf{Q}}$ exhibits a linear dependence on $(|\mathbf{Q}|/k_F)^2$~\cite{Suppl}, consistent with  the observation in Ref.~\onlinecite{Ries2015}, which also applied it as a temperature gauge for the experiment. It is particularly interesting to note the behavior of the peak at $\mathbf{Q}=0$ as $T$ is lowered through $T_{\text{BKT}}$. In the ``unitary'' regime [panel (b)], the two lowest temperatures are both below $T_{\text{BKT}}$, while the third, $T/T_F=0.1875 $, is above but close to it, as seen in Fig.~\ref{fig:BKTPhaseDiagram}. In comparison, in the BEC regime [panel (a)] only $T/T_F=0.0625 $ is below $T_{\text{BKT}}$, while  $T/T_F=0.125 $ is above but close to it. We see that the behavior of the peaks at $\mathbf{Q}=0$ in these systems shows a direct relation to where they are with respect to the transition temperature.

The fermion  momentum distribution is shown for the same systems in the insets of Fig.~\ref{fig:MomentDistrit}. In contrast with the pair momentum distribution, $n(\mathbf{k})$ shows significantly less temperature dependence. In the weakly interacting BCS regime, we see the steplike function around the Fermi surface at low $T$, as expected. As $T/T_F$ is increased, more fermions become thermally excited, with $n(\mathbf{k})$ showing substantial modification from the ground-state result at the highest $T$ shown, which is approximately $20\,\times\,T_{\text{BKT}}$. As the interaction strength is increased to the other two values, $n(\mathbf{k})$ is increasingly broader due to interaction effects reflecting the BCS-BEC crossover. However, its response to temperature variation becomes much reduced, and is barely noticeable in the BEC regime. The $n(\mathbf{k})$ results at the lowest temperature is in close agreement with the ground-state results~\cite{Shihao2015}. 

\begin{figure}[b]
\centering
\includegraphics[width=0.98\columnwidth]{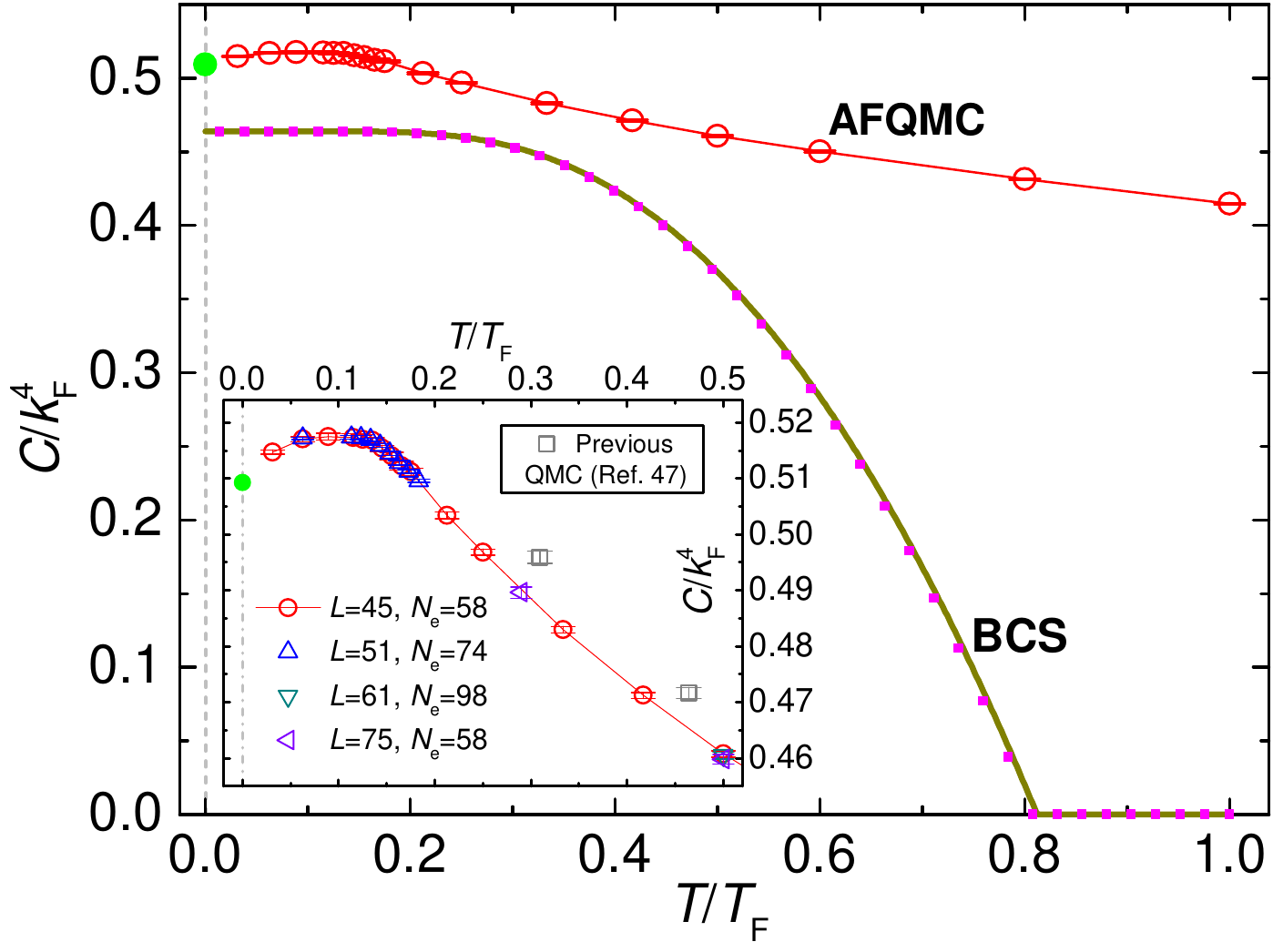}
\caption{\label{fig:CrossoverContact} The contact density $C/k_F^4$ as a function of temperature in the crossover regime [$\log(k_Fa)=+0.50$]. The red open circles represent the $T>0$ results from our AFQMC calculations. Results at ground state (green solid circle, from Ref.~\onlinecite{Shihao2015}) and from the BCS mean-field theory (dark yellow line for the continuum limit, magenta solid squares for $L=45,N_e=58$) are also included for comparisons. The inset is an enlargement of the contact density in $T/T_F\in[0,0.5]$ containing our AFQMC results from different systems up to $L=75$, and the previous lattice QMC results with $L=19$~\cite{Anderson2015}.}
\end{figure}

The contact $\mathcal{C}$~\cite{Tan2008a,Tan2008b,Tan2008c} is an important quantity in the strongly interacting Fermi gas, and it governs the asymptotic behaviors of several key properties in momentum space, for example, $n(\mathbf{k})$. In the 3D unitary Fermi gas, experimental measurements of the contact across the superfluid transition~\cite{Sagi2012,Wild2012,Carcy2019,Mukherjee2019} have allowed a thorough comparisons with various numerical results~\cite{Jensen2020}. In the 2D Fermi gas, the contact was experimentally measured~\cite{Giamarchi2012} at $T/T_F=0.27$ and numerically calculated via ground state QMC methods~\cite{Bertaina2011,Shihao2015}. The contact as a function of temperature was also studied~\cite{Anderson2015}, though this was limited to the normal phase and rather small system size as mentioned earlier.

Here we report exact numerical results of the contact  in the full range of temperatures crossing the BKT transition, in large lattice sizes, for the strongly interacting 2D Fermi gas. We compute the contact density $C=\mathcal{C}/N_s$ in units of $k_F^4$ via the double occupancy $D$ as $C/k_F^4=m^2U^2D/(4\pi^2 n^2)$. We have also confirmed that the asymptotic behavior of $n(\mathbf{k})k^4\sim\mathcal{C}$ at mediate to low temperatures and extracted the contact from the tail fitting of $n(\mathbf{k})$, which yielded consistent results for $\mathcal{C}$~\cite{Suppl}. Our results of the contact are shown in Fig.~\ref{fig:CrossoverContact}. While the BCS mean-field theory predicts a phase transition at around $T/T_F\simeq 0.8$ ($C_{\text{BCS}}/k_F^4$ is proportional to the square of the mean-field order parameter), our QMC results show an increase of the $C/k_F^4$ as the temperature is lowered, followed by a shallow maximum around the BKT transition, and then a decrease which smoothly connects with the ground state result~\cite{Shihao2015}. This behavior is qualitatively different from 3D, where the contact shows a dramatic increase when entering the superfluid phase~\cite{Carcy2019,Mukherjee2019,Jensen2020}. 

In summary, employing major advances in AFQMC algorithms, we have studied the finite-temperature properties of 2D Fermi gas with zero-range attractive interaction. Reaching large lattice sizes, we scan a wide range of interaction strengths and temperatures, and determine the phase diagram of the BKT transition. We systematically characterize the BCS-BEC crossover by the pairing wave functions in both reciprocal and real space. We compute both the fermion and pair momentum distributions at finite temperature, and observe behaviors consistent with experimental results. We have also accurately determined the contact versus temperature, and find that it exhibits different behaviors from the 3D case which has been well characterized experimentally.

We hope that these results will serve as a useful guide for experiments, and provide comparison and benchmark for the many analytical and computational studies being stimulated by the intense ongoing experimental efforts. This study also paves the way for further precision many-body computations in the 2D Fermi gas, including effective range effects~\cite{Mulkerin2020a,Mulkerin2019, Mulkerin2020b}, the pseudogap phenomena~\cite{Feld2011,Matsumoto2017,Mulkerin2020a}, and spin-orbit coupling~\cite{Shihao2016PRL}, among many others.

We thank Qijin Chen, Xingcan Yao and Wei Zhang for helpful discussions. Y. Y. H. thanks the hospitality of Renmin University of China and Institute of Physics, Chinese Academy of Sciences, where part of this work was done during the academic visits. Y. Y. H. acknowledges the National Natural Science Foundation of China (NSFC) under Grant No. 12047502. The Flatiron Institute is a division of the Simons Foundation.

\bibliography{2DFermiGasMain}


\newpage
\onecolumngrid

\begin{center}
\textbf{\large Supplementary material for ``Precision Many-Body Study of the Berezinskii-Kosterlitz-Thouless Transition and Temperature-Dependent Properties in the Two-Dimensional Fermi Gas''}
\end{center}

\section{I. The condensate fraction and BKT transition temperatures}

\hspace{0.5cm} In this section, we first present our numerical results of the BKT transition temperatures. Then in the rest of this section, we present the details on how we extract the BKT transition temperatures from the condensate fraction as a function of temperature, as well as the careful finite-size and finite-$N_e$ scaling of the transition temperature to reach the dilute gas limit. 

\begin{figure}[h!]
\centering
\includegraphics[width=0.55\columnwidth]{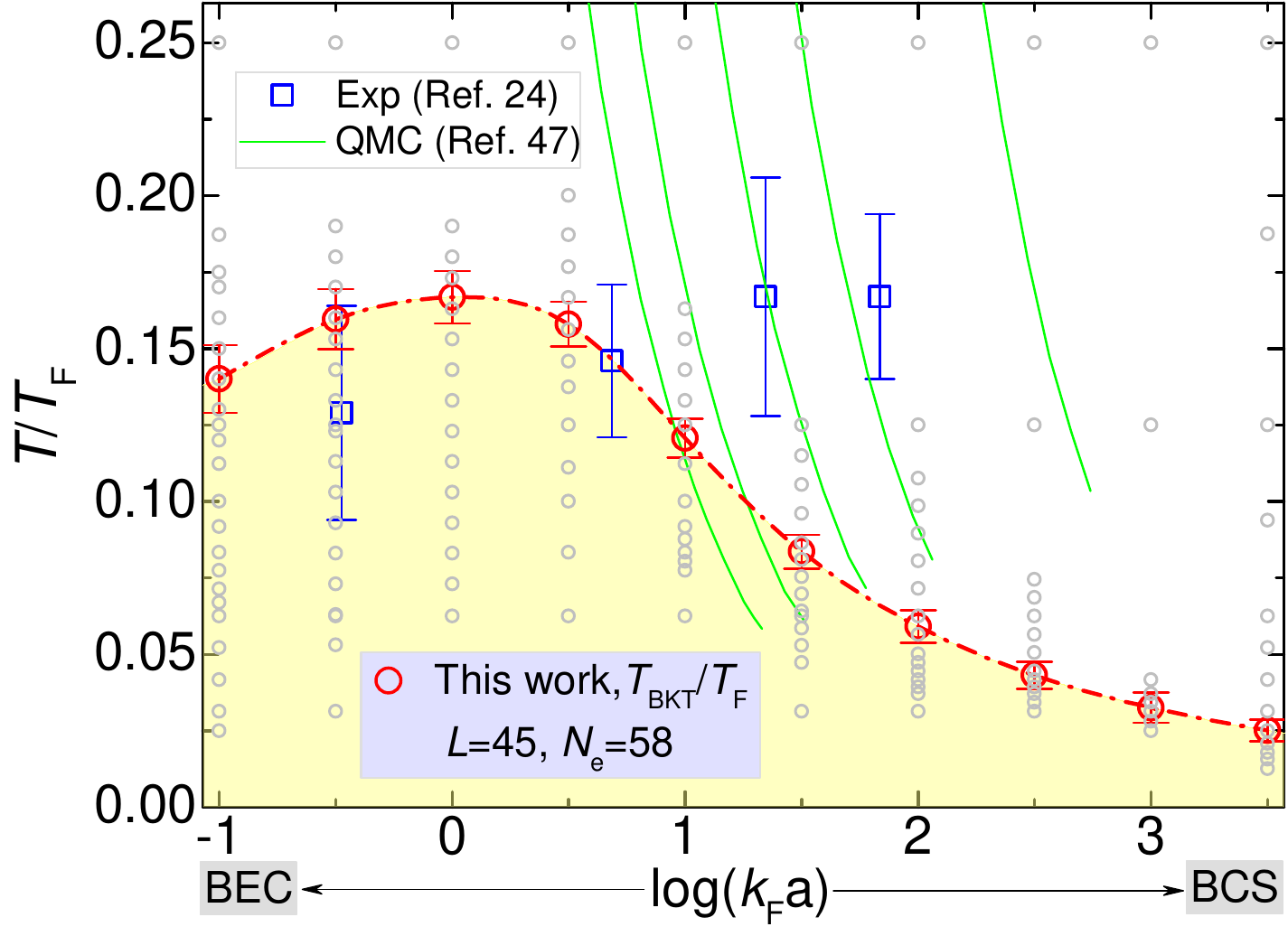}
\caption{\label{fig:OurTBKT} Illustration of the QMC calculations performed in this and prior work. Empty Red circles represent the BKT transition temperatures calculated in this work for $L=45,N_e=58$ with quadratic dispersion. The gray, hollow circles identify where we have performed AFQMC simulations. Previous lattice QMC simulations~\cite{Anderson2015} were performed along the green paths. The BKT transition temperatures measured experimentally~\cite{Ries2015} are also shown (blue, hollow squares).}
\end{figure}

\hspace{0.5cm} In Fig.~\ref{fig:OurTBKT}, we show our QMC results of the BKT transition temperatures, $T_{\text{BKT}}/T_F$, which are obtained in this work from an extensive set of individual AFQMC calculations marked by gray circles. The most accurate finite-temperature results have been provided by lattice QMC calculations~\cite{Anderson2015}, which were limited mostly to higher temperatures in the normal phase, as seen from the parameter paths shown as green lines. The experimental measurements~\cite{Ries2015} are also included.

\hspace{0.5cm} The BKT transition temperature for a two-dimensional system is usually determined by finite-size scaling, such as in previous studies for attractive Hubbard model~\cite{Moreo1991,Santos1993,Paiva2004,Nakano2006}. However, the results always showed large finite-size effects~\cite{Paiva2004}, and it was demonstrated that one might need to reach linear system size $L\sim 100$ to recover the critical exponent $\eta_c=0.25$ for the real-space correlation function~\cite{Shailesh2002}. Such a finite-size scaling is even more difficult for the 2D Fermi gas system, since the two-step convergence -- first of system size and then of the number of fermions - to the continuum and bulk limits is hard to achieve together. In this work we turn to another way via the condensate fraction to determine the BKT transition temperatures. 

\begin{figure}[h!]
\centering
\includegraphics[width=0.85\columnwidth]{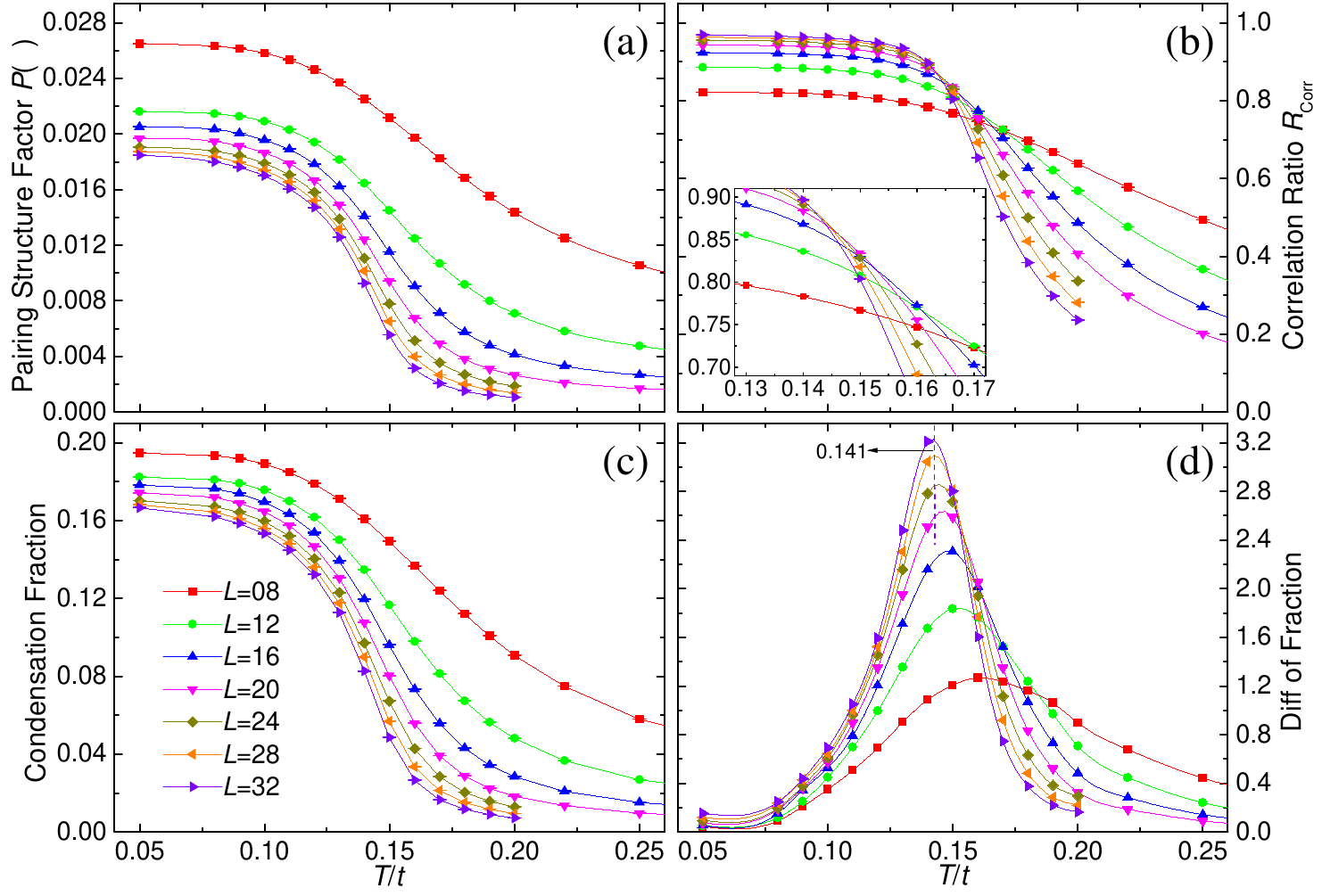}
\caption{\label{fig:HubbardTest} Determining the BKT transition temperature in the 2D attractive Hubbard model.The temperature dependence of (a) the pairing structure factor $P(\boldsymbol{\Gamma})$, (b) the correlation ratio $R_{\text{Corr}}$, (c) the condensate fraction and (d) first-order (numerical) derivative of the condensate fraction are plotted. The inset of (b) shows the convergence of the crossing points of finite-size results for $R_{\text{Corr}}$. The system parameters are $U/t=-4,\mu/t=0.6$, and results are presented for linear lattice size $L=8,12,16,20,24,28,32$. }
\end{figure}

\hspace{0.5cm} To demonstrate the validity of this method, we first present benchmark results on the 2D lattice Hubbard model with attractive interaction. We choose the model parameters $U/t=-4,\mu/t=0.6$ (where $t$ is the nearest-neighbor hopping constant), for which the filling of the system is approximately 0.575 at the low temperature regime. To make a comparison with previous study~\cite{Paiva2004}, we first compute the structure factor for on-site spin-singlet pairing as $P(\mathbf{k})=\frac{1}{N_s^2}\sum_{\mathbf{ij}}e^{i\mathbf{k}\cdot(\mathbf{R}_{\mathbf{i}}-\mathbf{R}_{\mathbf{j}})}\langle\hat{\Delta}_{\mathbf{i}}^+\hat{\Delta}_{\mathbf{j}}+\hat{\Delta}_{\mathbf{i}}\hat{\Delta}_{\mathbf{j}}^+\rangle/2$ with $\hat{\Delta}_{\mathbf{i}}^+=c_{\mathbf{i}\uparrow}^+c_{\mathbf{i}\downarrow}^+$ ($N_s=L^2$ as the number of lattice sites of the 2D system with $L$ denoting the linear system size) at $\mathbf{k}=\boldsymbol{\Gamma}$ point, and the corresponding correlation ratio $R_{\text{Corr}}=1-P(\boldsymbol{\Gamma}+\mathbf{q})/P(\boldsymbol{\Gamma})$, where $\mathbf{q}$ is the smallest momentum on the lattice, i.e., $(0,2\pi/L)$ or $(2\pi/L,0)$. As shown in Fig.~\ref{fig:HubbardTest}(b), the crossing points  of $R_{\text{Corr}}$ converge to as $L$ is increased, yielding a location of the BKT transition around $T/t\simeq 0.142$, which is consistent with the result in Ref.~\onlinecite{Paiva2004}. 

\hspace{0.5cm} We then compute the condensate fraction in this model, and its numerical derivative with respect to the temperature, following the description in the main text, and present the results in Fig.~\ref{fig:HubbardTest}(c)(d). We observe sharp peaks in the derivatives of condensate fraction, and their location to a position consistent with the BKT transition temperature obtained from the correlation ratio as demonstrated in Fig.~\ref{fig:HubbardTest}(b). Although in the thermodynamic limit the condensate fraction vanishes except at $T=0$\,K, it has a finite value in a finite system as defined via the pairing matrix (main text). The condensate fraction is intimately connected to the first-order correlation function which has a qualitative change of behavior at the BKT transition, from exponential to power-law decay. As $T$ falls below $T_{\rm BKT}$,  the condensate fraction exhibits a rapid increase, as observed in Fig.~\ref{fig:HubbardTest}(c), and its derivative shows a peak which becomes sharper as system size is increased, as seen in Fig.~\ref{fig:HubbardTest}(d).

\begin{figure}[h!]
\centering
\includegraphics[width=0.49\columnwidth]{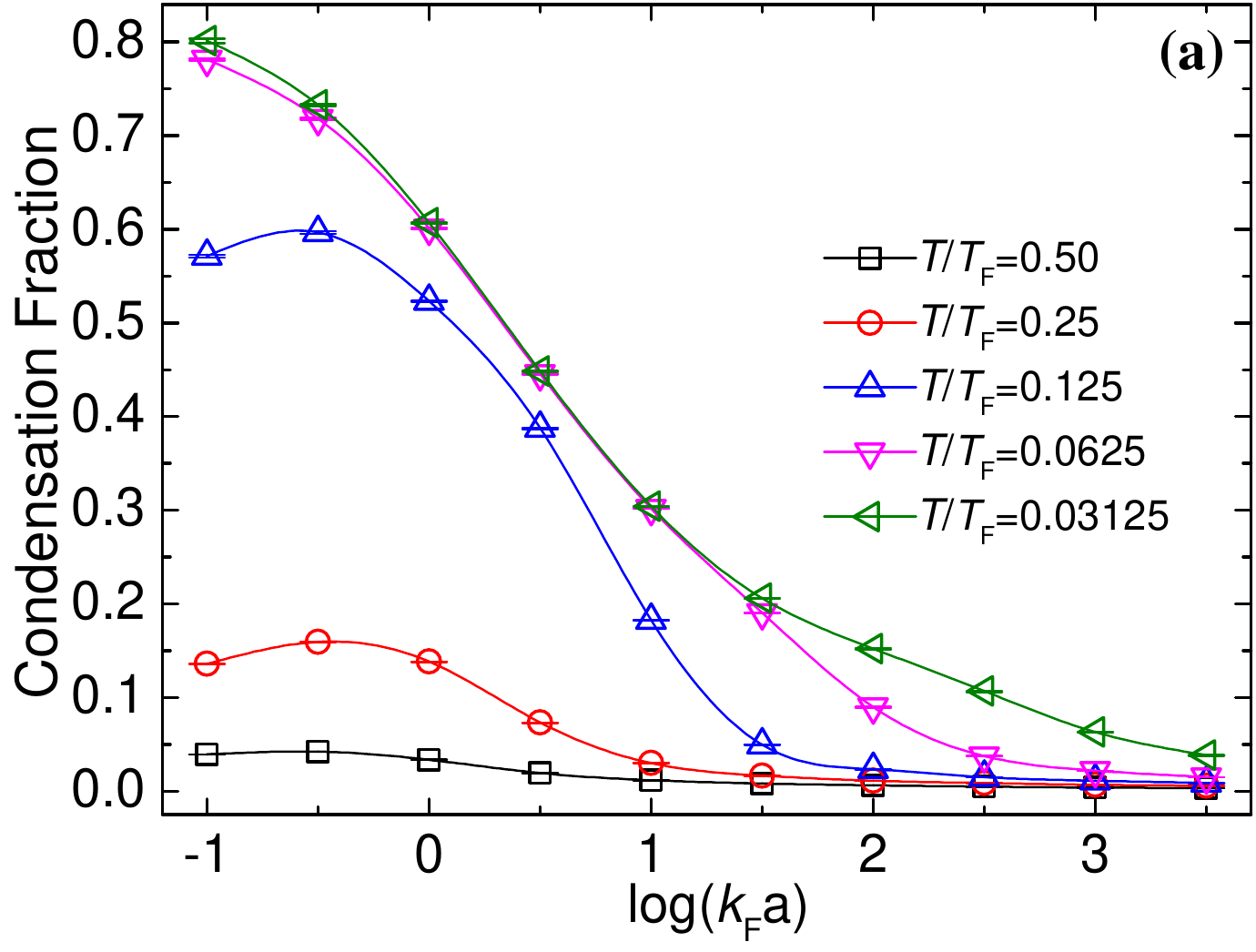}
\includegraphics[width=0.49\columnwidth]{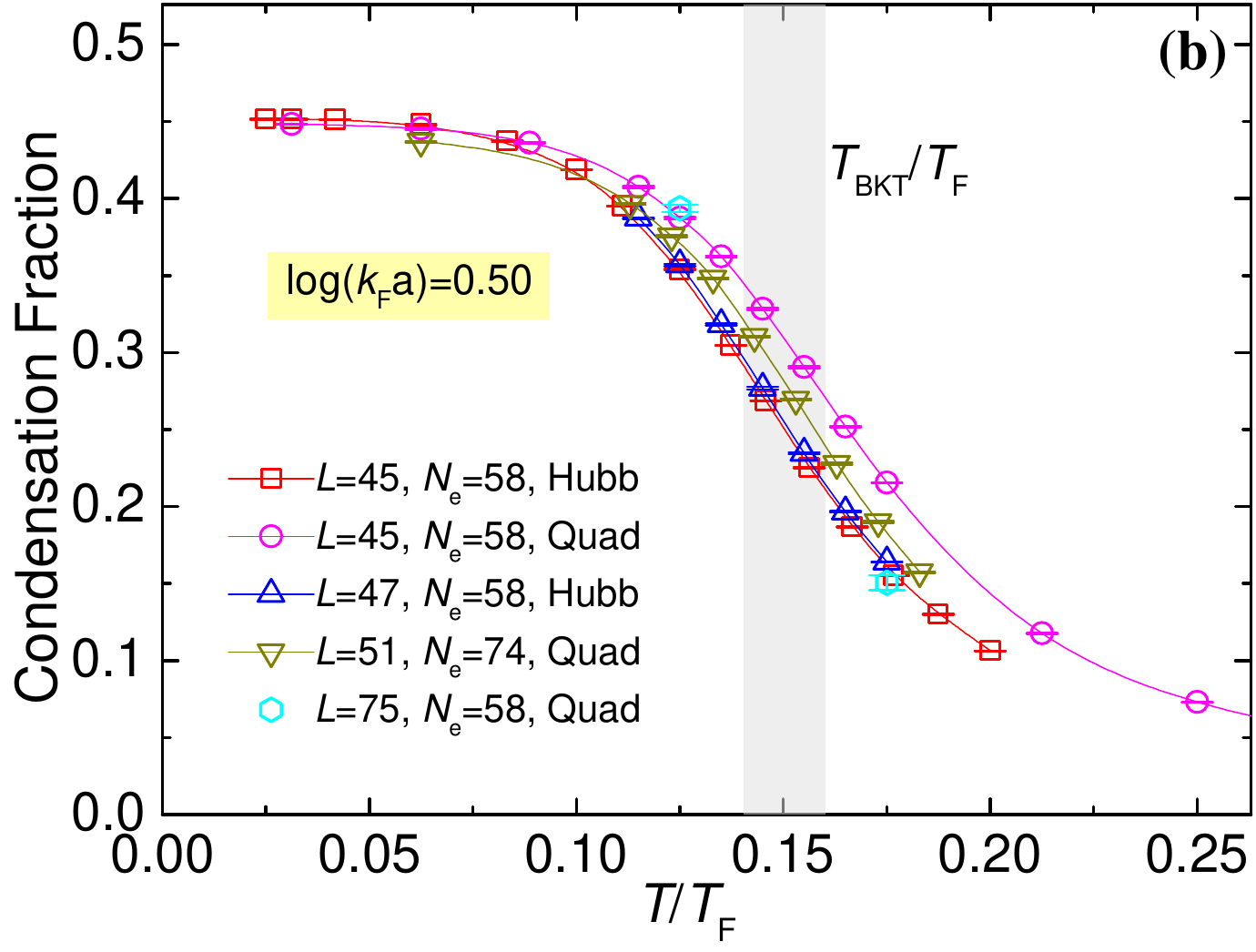}
\caption{\label{fig:CondnstFrac} AFQMC results of the condensate fraction, (a) as a function of $\log(k_Fa)$ at various temperatures $T/T_F=0.50 \sim 0.0325$ for the system $L=45,N_e=58$ with the quadratic dispersion; (b) as a function of temperature for $\log(k_Fa)=+0.50$ with several different system sizes and both Hubbard and quadratic dispersions (termed as "Hubb" and "Quad" respectively). }
\end{figure}

\begin{figure}[h!]
\centering
\includegraphics[width=0.55\columnwidth]{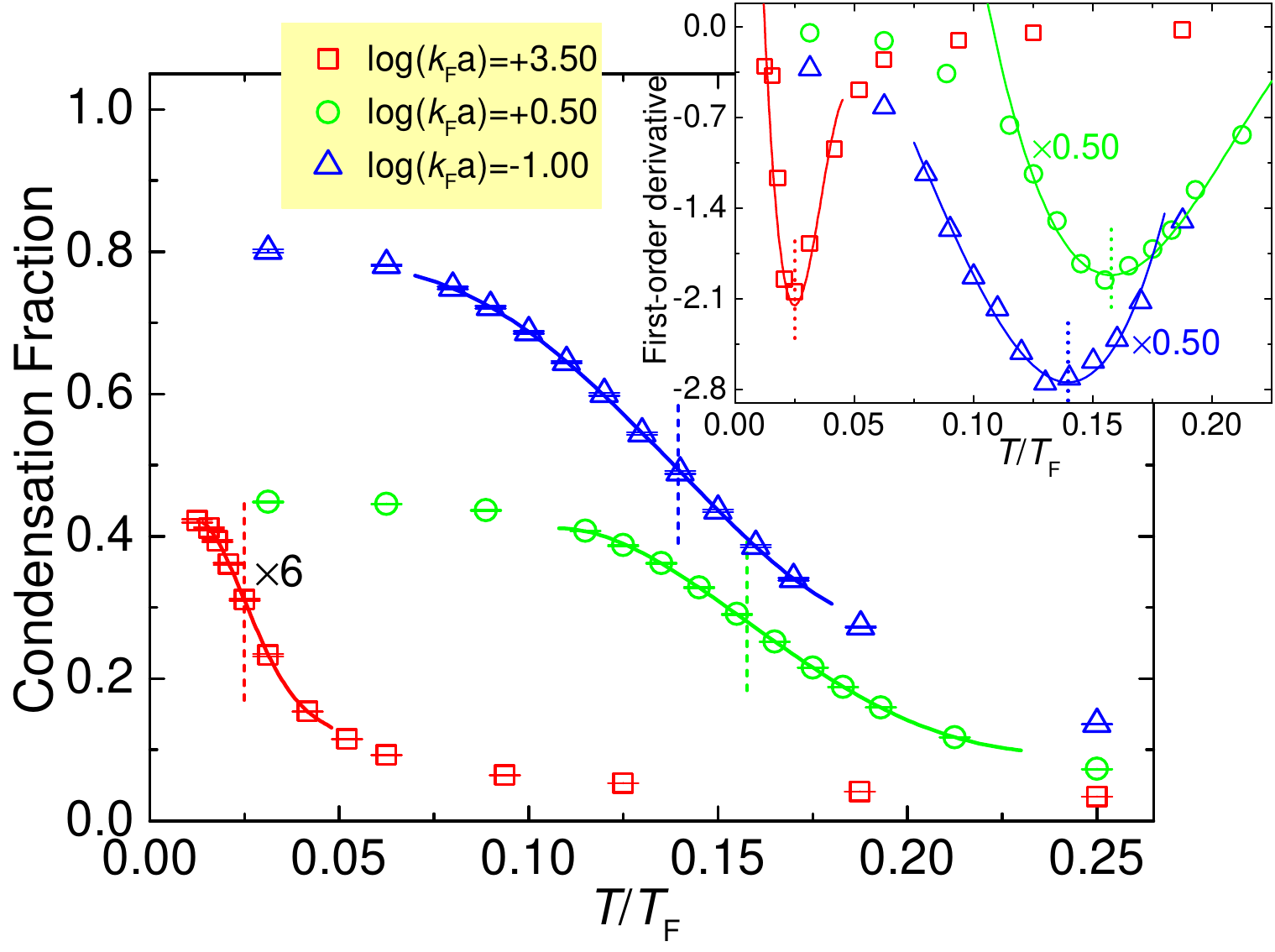}
\caption{\label{fig:Derivative} Obtaining the BKT transition temperatures from the condensate fraction, for $\log(k_Fa)=+3.5,+0.5,-1.0$. The AFQMC data points are obtained on $L=45,N_e=58$ with quadratic dispersion. The solid lines show the fourth-order polynomial fit around the transition points. The inset shows the first-order derivative of condensate fractions, obtained from both numerical finite differences and the analytical formula of the fitted fourth-order polynomials. [Note that we have scaled the original data of the condensate fraction at $\log(k_Fa)=+3.50$ by a factor of $\times6$ in the main plot, and the derivatives fat $\log(k_Fa)=+0.50$ and $\log(k_Fa)=-1.00$ by $\times 0.50$ respectively in the inset, to fit into the plots. }
\end{figure}

\hspace{0.5cm} Turning to the 2D Fermi gas, we first present condensate fraction data for a variety of finite-size systems for reference. At a fixed $\log(k_Fa)$ we scan a wide range of temperatures. In Fig.~\ref{fig:CondnstFrac}(a), we present the results of condensate fraction for $\log(k_Fa)\in[-1.0,+3.5]$ and five characteristic temperatures, for system size $L=45,N_e=58$. The condensate fraction is small at high temperatures, as expected for the normal phase. At the lowest temperature $T/T_F=0.03125$, the result increases monotonically from from the BCS regime to thee BEC regime, and is very close to the corresponding ground state result in Ref.~\onlinecite{Shihao2015}. At intermediate temperatures, the condensate fraction shows a broad peak in the crossover regime, which is consistent with the final result of the BKT transition temperature (Fig.~1 of the main text), which reaches a maximum in the crossover regime. In Fig.~\ref{fig:CondnstFrac}(b), we plot the condensate fraction in the crossover regime ($\log(k_Fa)=+0.50$) as a function of temperature for several different system sizes, and two different forms of the dispersions. We see that the results are very consistent for each physical system (fixed $N_e$ and interaction). Typically, the Hubbard dispersion can mimic the quadratic dispersion under the continuum limit (fixed number of fermions $N_e$ with the fermion filling $n=N_e/N_s$ approaching zero), where they should present the same results for physical observables. However, they can have substantially different finite-size effect, especially going towards the BEC side with increasing interaction strength. As a result, the BKT transition temperatures, obtained from systems with the same $L$ and $N_e$ but different kinetic dispersions, can be different. We find that this difference can be non-negligible going to the BEC side, while it approaches zero statistically towards the BCS side. We have also identified that the results from quadratic dispersion have smaller finite-size effect and cleaner finite-size scaling behavior than those from Hubbard dispersion. Thus, in the following, we concentrate on the results obtained from quadratic dispersion. 

\begin{figure}[h!]
\centering
\includegraphics[width=0.32\columnwidth]{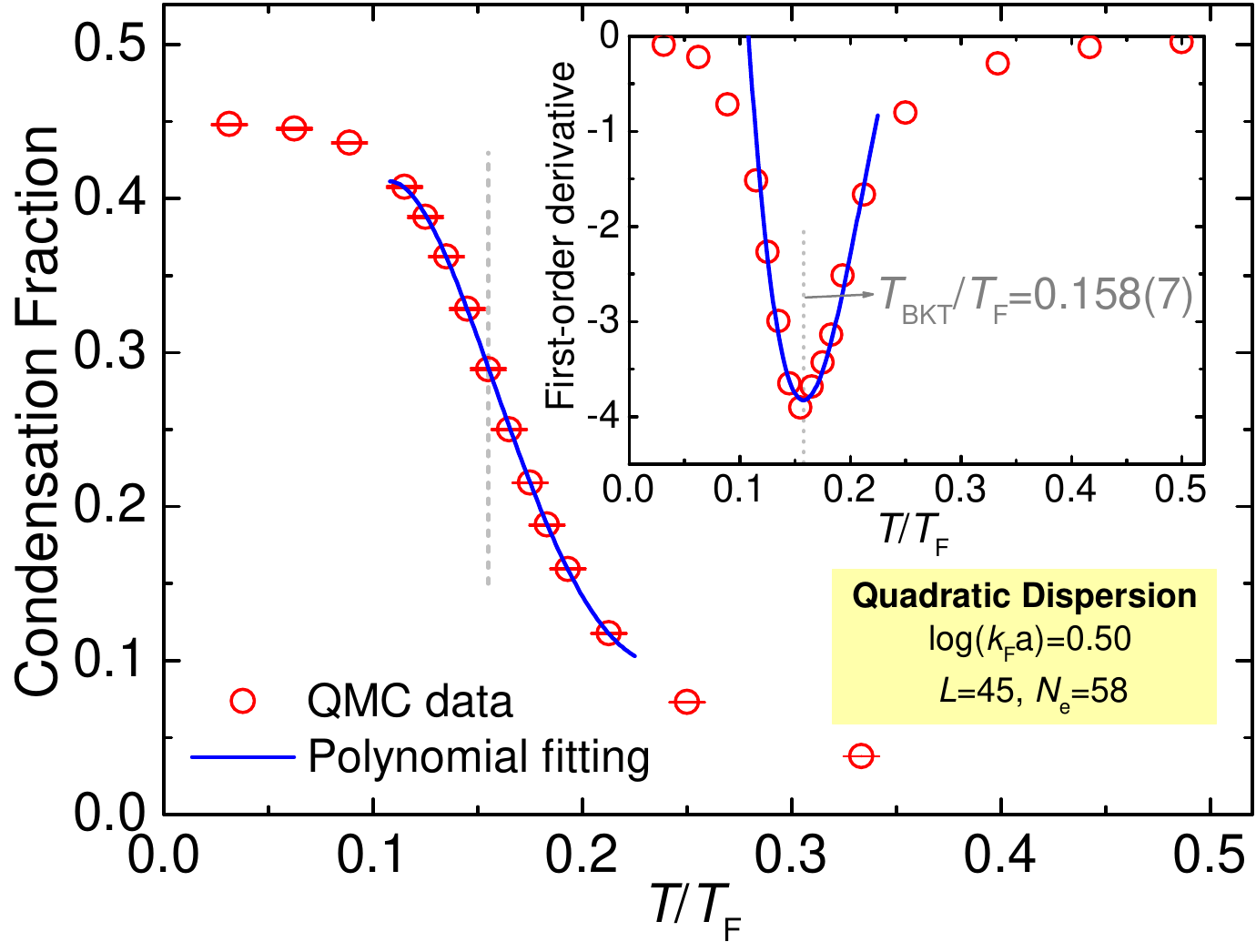}
\includegraphics[width=0.32\columnwidth]{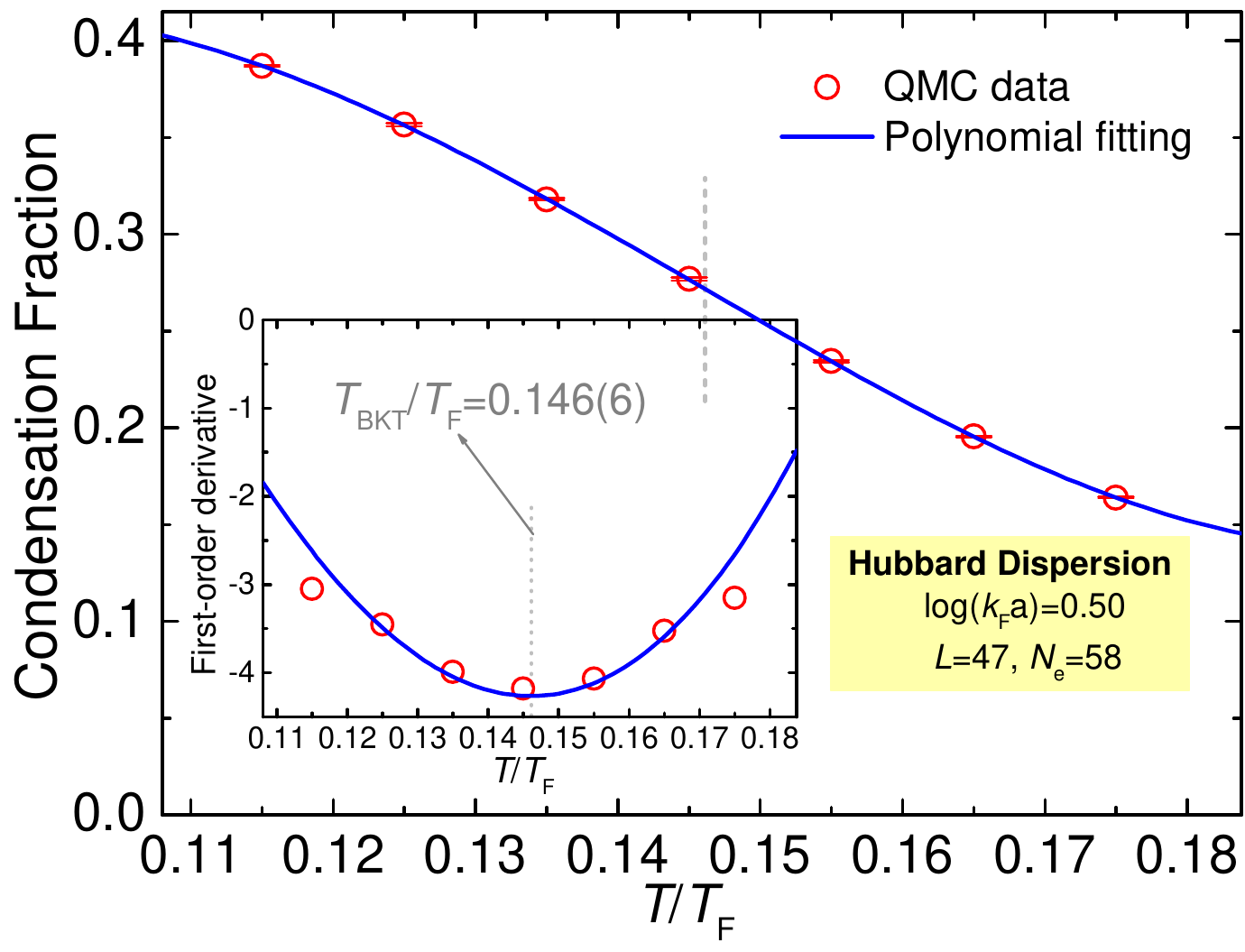}
\includegraphics[width=0.32\columnwidth]{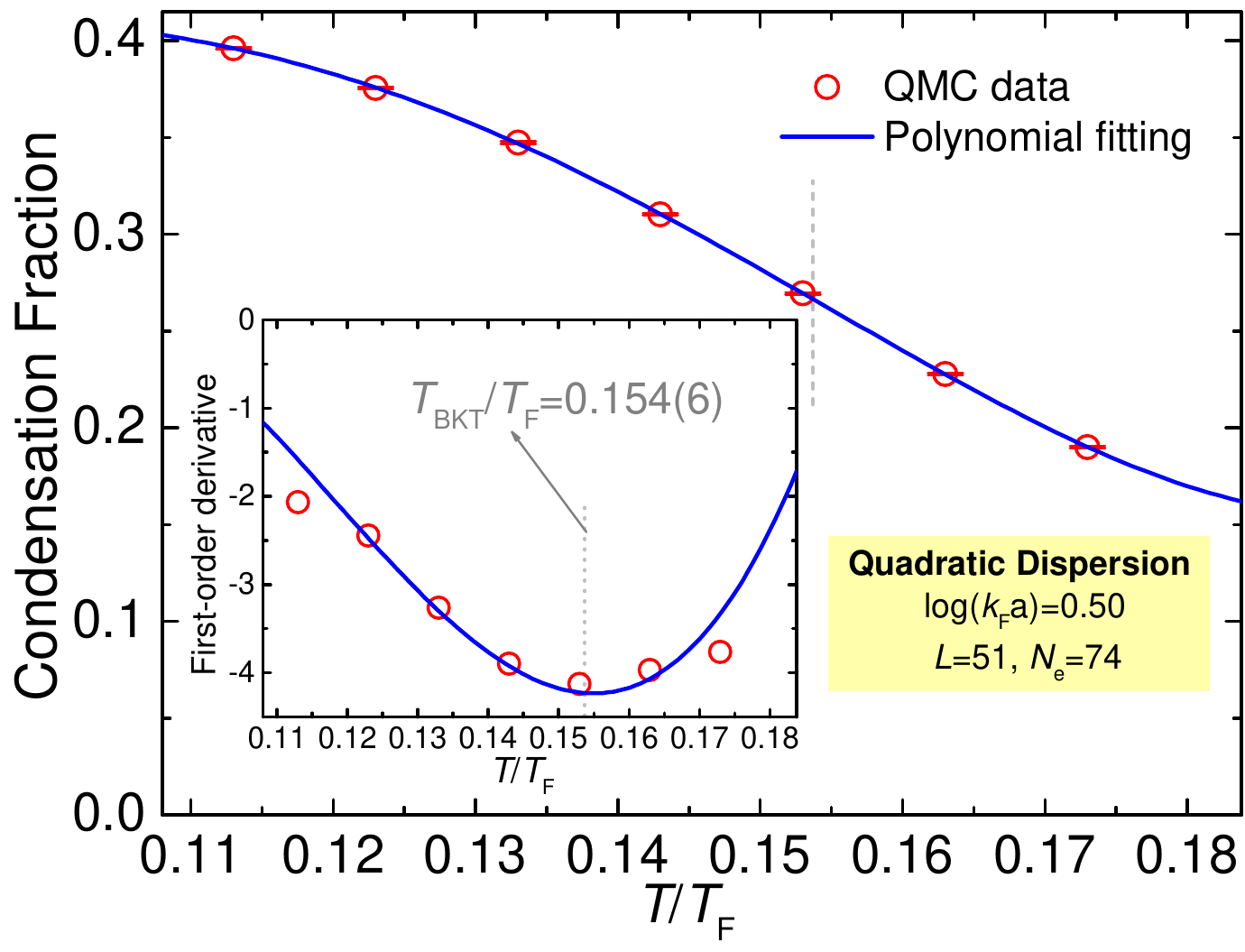}
\caption{\label{fig:Convergence} Obtaining the BKT transition temperatures from the condensate fraction, for $\log(k_Fa)=+0.5$. The BKT transition temperatures is calculated in three different systems as: $L=45,N_e=58$ with quadratic dispersion, $L=47,N_e=58$ with Hubbard dispersion, $L=51,N_e=74$ with quadratic dispersion. The results of the condensate fraction of all these systems are shown in Fig.~\ref{fig:CondnstFrac}(b). The calculated BKT transition temperatures are indicated in the insets.}
\end{figure}

\begin{figure}[h!]
\centering
\includegraphics[width=0.99\columnwidth]{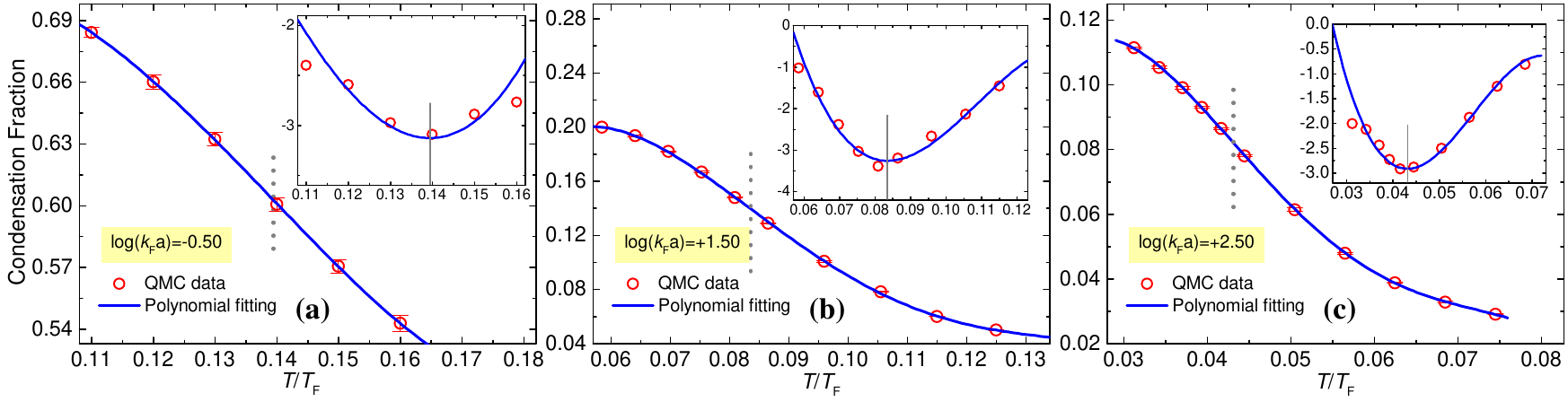}
\caption{\label{fig:m050p150p250} The condensate fraction and BKT transition temperatures at $\log(k_Fa)=-0.5$ ($L=57,N_e=42$) and $\log(k_Fa)=+1.5,+2.5$ ($L=45,N_e=58$), with quadratic dispersion. Symbols and layouts are the same as in Fig.~\ref{fig:Convergence}. The obtained BKT transition temperatures are $T_{\text{BKT}}/T_F=0.139(9)$ for $\log(k_Fa)=-0.50$, $T_{\text{BKT}}/T_F=0.084(6)$ for $\log(k_Fa)=+1.50$, and $T_{\text{BKT}}/T_F=0.043(4)$ for $\log(k_Fa)=+2.50$.}
\end{figure}

\hspace{0.5cm} We next apply the idea benchmarked above in the attractive Hubbard model to the 2D Fermi gas, to determine the BKT transition temperature at each $\log(k_Fa)$. We first carry out AFQMC simulations with a coarse grid of temperatures to find the approximate location of the peak in the first-order derivative of the condensate fraction. Then we perform AFQMC simulations with a finer temperature grid ($\sim \Delta T/T_F=0.01$) around that peak location, to more precisely determine the BKT transition temperature. To minimize the bias from the finite temperature grid in the numerical derivative, we use a fourth-order polynomial to fit the data points around the transition. The BKT transition temperature is given as the peak location of the derivative of the fourth-order polynomial. 

\hspace{0.5cm} In Fig.~\ref{fig:Derivative}, we present three examples of the procedure above, at three typical interaction strengths, $\log(k_Fa)=-1.00$ in the BEC regime, $\log(k_Fa)=+0.50$ in the crossover regime and $\log(k_Fa)=+3.50$ in the BCS regime. In Fig.~\ref{fig:Convergence}, we illustrate the finite-size and finite-$N_e$ effects, as well as the effect of using different forms of the kinetic energy dispersion. We compute the BKT transition temperatures for three different systems at $\log(k_Fa)=+0.50$ (in the crossover regime), with different system size, $N_e$ as well as kinetic energy dispersion. The results are consistent with each other and with that shown in Fig.~\ref{fig:Derivative}. To further reduce the finite-size effect towards the BEC side, we performed simulations in $L=57,N_e=42$ systems with quadratic dispersion for cross-checks. In Fig.~\ref{fig:m050p150p250}, we further show the results for three other interaction strengths, $\log(k_Fa)=-0.50,+1.50,+2.50$. The results of BKT transition temperatures demonstrated in Fig.~\ref{fig:Derivative} and Fig.~\ref{fig:m050p150p250}, which are obtained from the finite-size system of $N_e=58$ with $L=45$, are then summarized in Fig.~\ref{fig:OurTBKT} and Fig.~1 in the main text. 

\begin{figure}[h!]
\centering
\includegraphics[width=0.99\columnwidth]{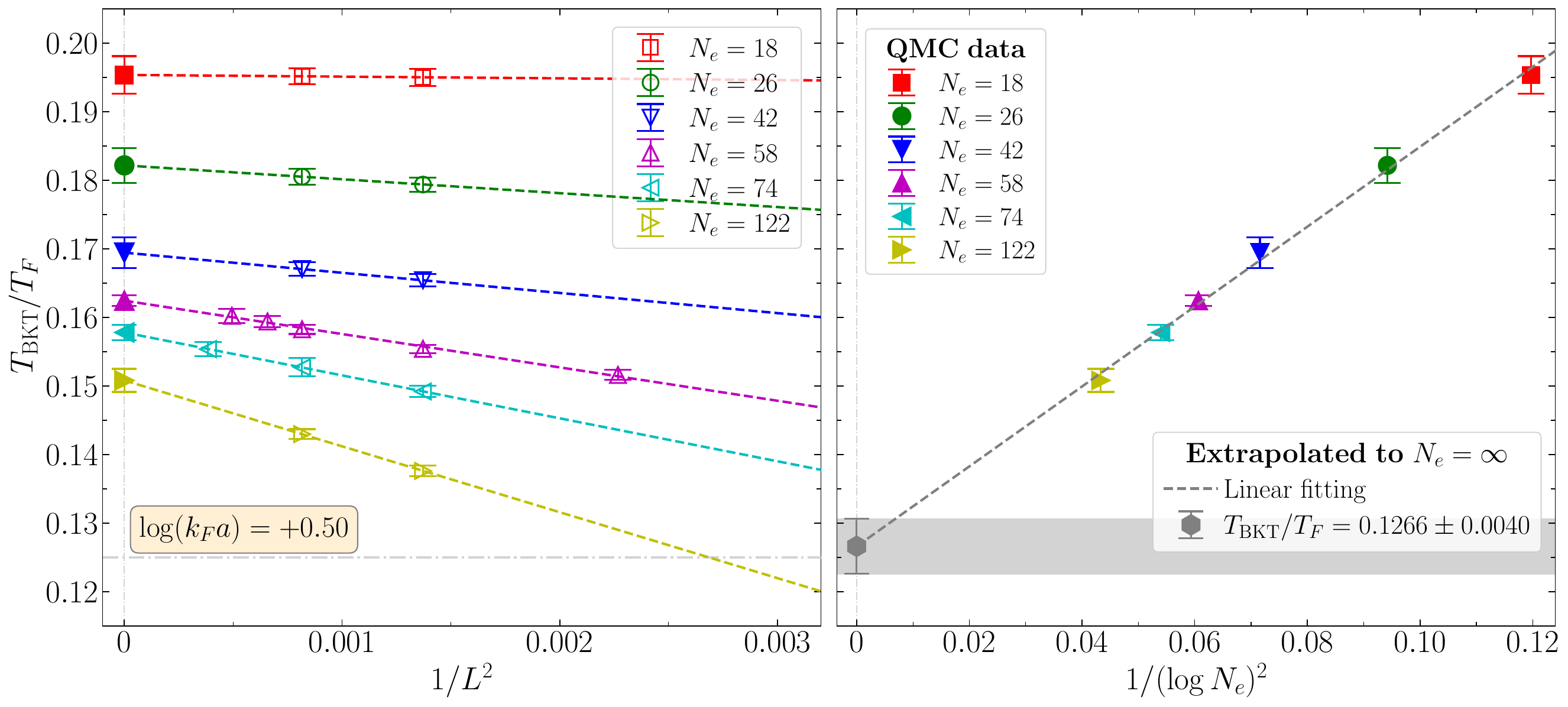}
\caption{\label{fig:FScaleLogkFa050} 
Illustration of the finite-size scaling of the BKT transition temperatures, at $\log(k_Fa)=+0.5$ with quadratic dispersion. The left panel shows the extrapolation to the continuum limit for fixed number of particles $N_e$, $T_{\rm BKT}(L,N_e)/T_F$ vs.~$1/L^2$. The right panel shows the finite-size scaling of the resulting $T_{\rm BKT}(L\rightarrow \infty,N_e)/T_F$ with $N_e$.
}
\end{figure}

\hspace{0.5cm} For all the quantities we have computed, we find that the dependence on lattice size $N_s$ is the leading finite-size effects, and the dependence on the number of particles $N_e$ is weak for $N_e\ge 58$. For the finite-size and finite-$N_e$ BKT transition temperature $T_{\rm BKT}(L,N_e)/T_F$, however, it turns out that both are important and a finite-$N_e$ scaling {\it after the extrapolation to the continuum limit\/} leads to non-negligible corrections on the $N_e=58$ results. This is illustrated in Fig.~\ref{fig:FScaleLogkFa050} for one of the most challenging cases, which occurs  in the crossover regime, for  $\log(k_Fa)=+0.5$. Individual simulations for $\hat H(N_s,N_e)$ (with $N_s=L^2$) as described above yield $T_{\rm BKT}(L,N_e)/T_F$. We find that, for a fixed $N_e$, this transition temperature is well described by a linear dependence on $1/L^2$, consistent with previous studies~\cite{Werner2012,Shihao2015,YuanYao2019}. Extrapolation  to $L\rightarrow \infty$ is performed for each fixed $N_e$ as shown in the left panel of Fig.~\ref{fig:FScaleLogkFa050}. Then the results at the continuum limit,  $T_{\rm BKT}(L=\infty,N_e)/T_F$, are found to follow $1/(\log N_e)^2$ linearly to an excellent degree, which is consistent with expectations of the BKT transition~\cite{Tomita2002,Filinov2010,Nguyen2019}. This step is shown in the right panel of Fig.~\ref{fig:FScaleLogkFa050}, and presents a final result of $T_{\rm BKT}/T_F\equiv T_{\rm BKT}(L=\infty,\infty)/T_F=0.126(8)$, taking twice the statistical error bar from the final fit for a conservative estimate of the statistical uncertainty in the two-step finite-size scaling. Similar procedures were performed for several other representative interaction strengths as shown in the main text in Fig.~1 by solid red circles. Based on these results and those obtained from the finite system of $L=45,N_e=58$, we reach the final answer of the BKT transition temperatures spanning the BCS-BEC crossover via interpolation, shown as the solid red line with the shaded band indicating statistical error bars in Fig.~1 of the main text. 

\section{II. Finite-size effects illustrated in the single-particle momentum distribution}

\begin{figure}[htb!]
\centering
\includegraphics[width=0.95\columnwidth]{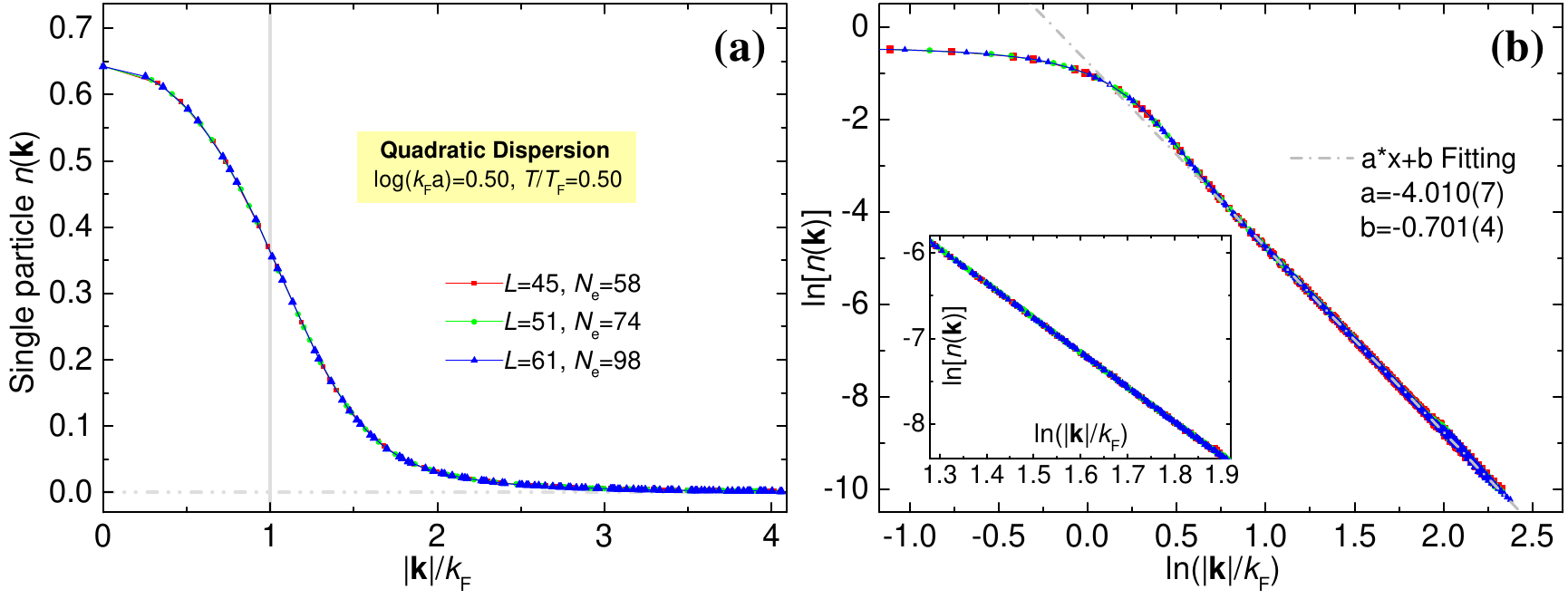}
\caption{\label{fig:ContactTOvTf050} The single-particle momentum distribution $n(\mathbf{k})$ for three different systems, $L=45,N_e=58$, $L=51,N_e=74$, and $L=61,N_e=98$, at $\log(k_Fa)=+0.50$ with $T/T_F=0.50$. The error bars are smaller than the symbol sizes. In (a), all the results colllapse to the same curve. In (b), $\log[n(\mathbf{k})]$ versus $\log(|\mathbf{k}|/k_F)$ are plotted, which is linear asymptotically. The slope value $-4.010(7)$ from the linear fit is consistent with the expected $1/|\mathbf{k}|^4$ behavior. }
\end{figure}

\begin{figure}[htb!]
\centering
\includegraphics[width=0.47\columnwidth]{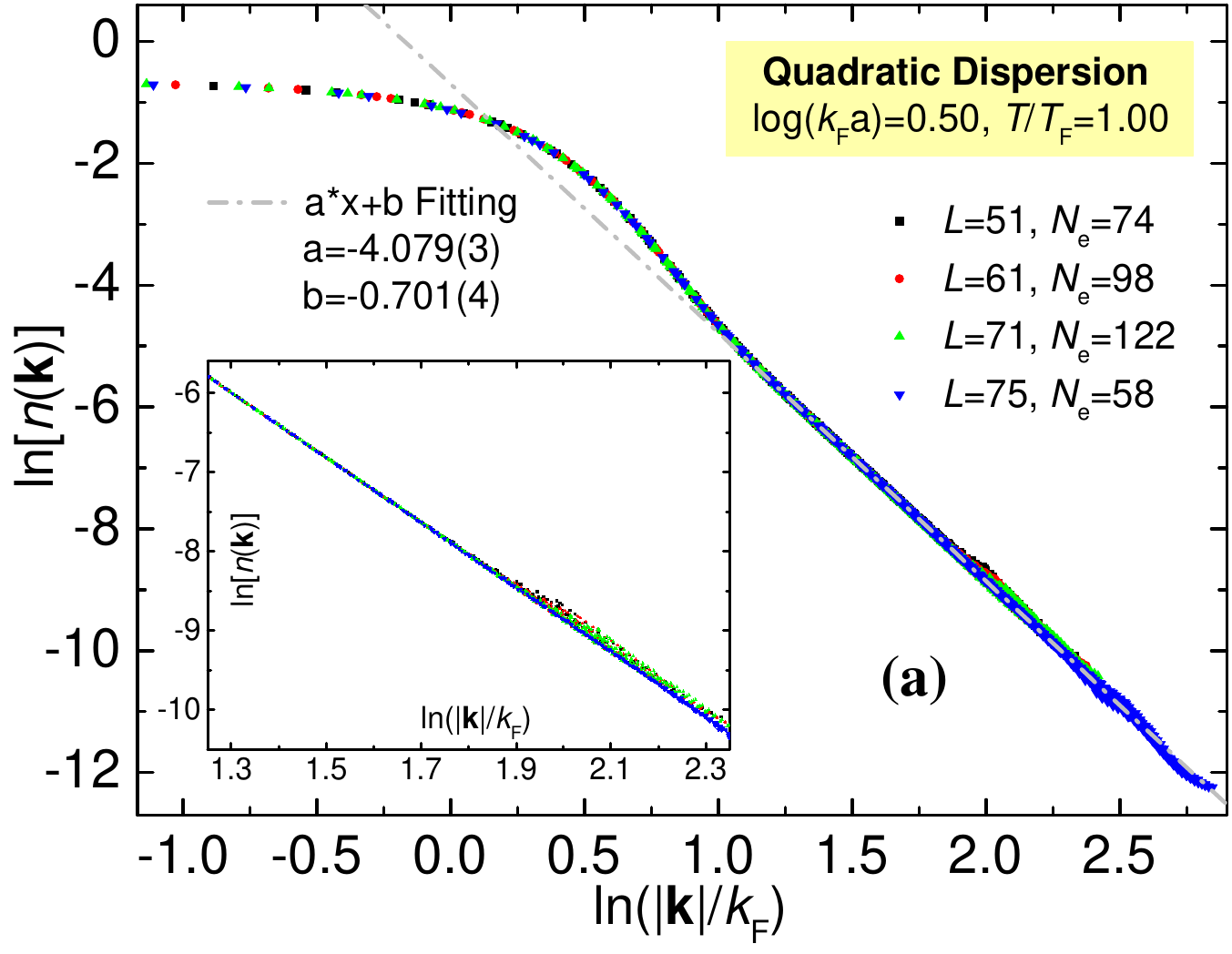} \hspace{0.1cm}
\includegraphics[width=0.47\columnwidth]{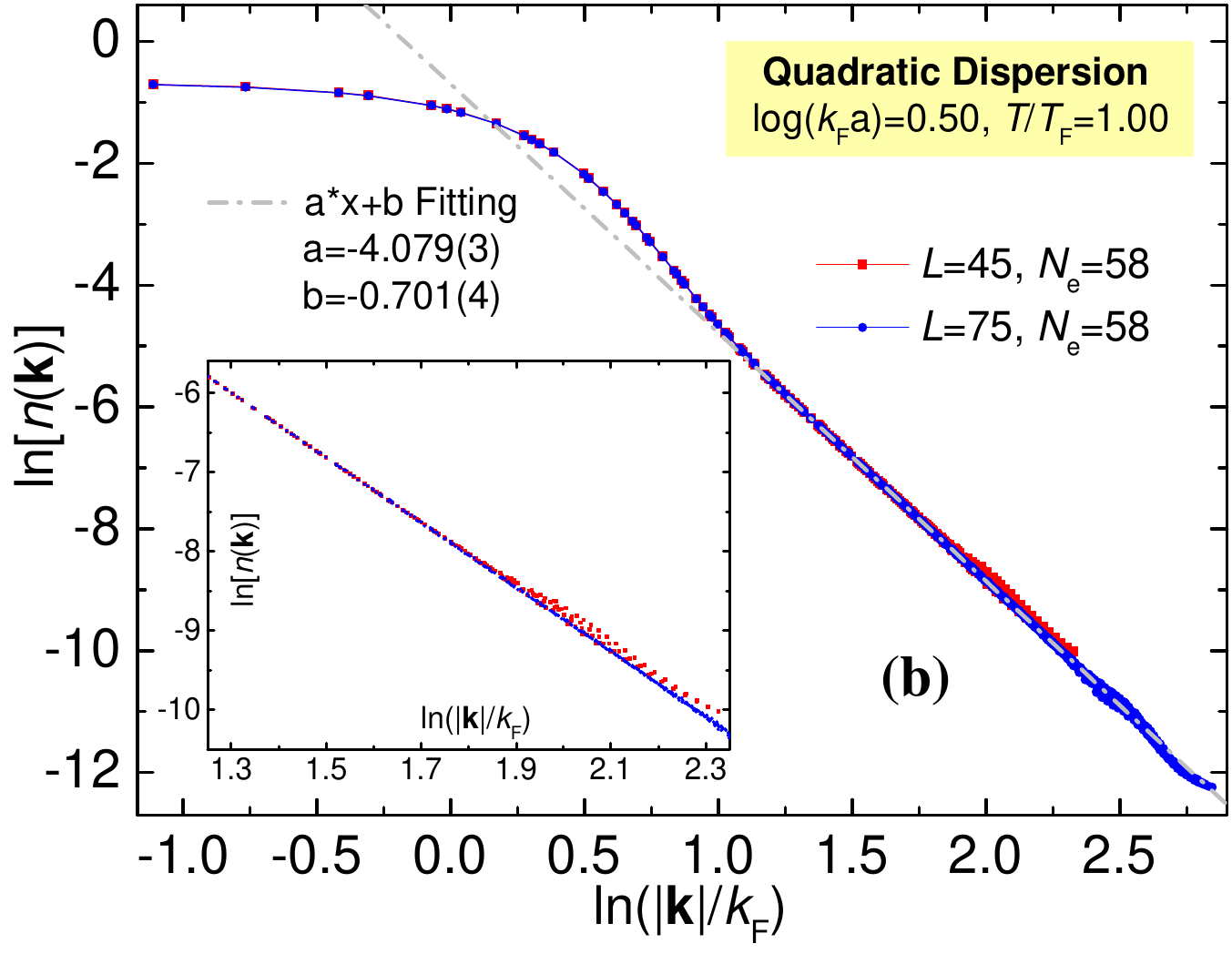}
\caption{\label{fig:ContactTOvTf100} 
Single-particle momentum distribution $n(\mathbf{k})$  for five different system sizes, at $\log(k_Fa)=+0.50$ with $T/T_F=1.00$.
Both panels are plotted on a log-log scale. The error bars are smaller than the symbol sizes. The best fit for $\log[n(\mathbf{k})]$ versus $\log(|\mathbf{k}|/k_F)$ gives a slope estimate of $-4.079(3)$.}
\end{figure} 

\hspace{0.5cm} We have performed large-scale simulations to high accuracy to quantify convergence towards the continuum and thermodynamic limit. In this section, we show further details on the single-particle momentum distribution $n(\mathbf{k})$ discussed in the main text. In Fig.~\ref{fig:ContactTOvTf050}, we present results at $T/T_F=0.50$ in  the crossover regime [$\log(k_Fa)=+0.50$].  Convergence to both the continuum limit and the thermodynamic limit is excellent, as seen from the collapse of the data for different $L$ (continuum) and different $N_e$. The fit of $\log[n(\mathbf{k})]$ versus $\log(|\mathbf{k}|/k_F)$ gives a slope $-4.010(7)$, which is consistent with the expected asymptotic behavior. Fig.~\ref{fig:ContactTOvTf100} shows the results at a higher temperature, $T/T_F=1.00$, for the same interaction strength. Here we do an even more extensive set of calculations to probe finite-size effects. The largest system size we explored is $L=75$ with $N_e=58$. Panel b examines the effect of going to the continuum limit with fixed number of particles, while panel a examines the effect of going to the bulk limit (and the interplay between lattice size and density). Again we see that the effect of finite-size in the calculations is negligible at the sizes we have reached. This slope from the linear fit of  $\log[n(\mathbf{k})]$ versus $\log(|\mathbf{k}|/k_F)$ is $-4.079(3)$ at $T/T_F=1.00$. Interestingly this value at higher $T$ shows a deviation from the theoretical prediction of $-4$. The deviation is small but seems robust against the different sizes in our simulations.

\begin{figure}[htb!]
\centering
\includegraphics[width=0.46\columnwidth]{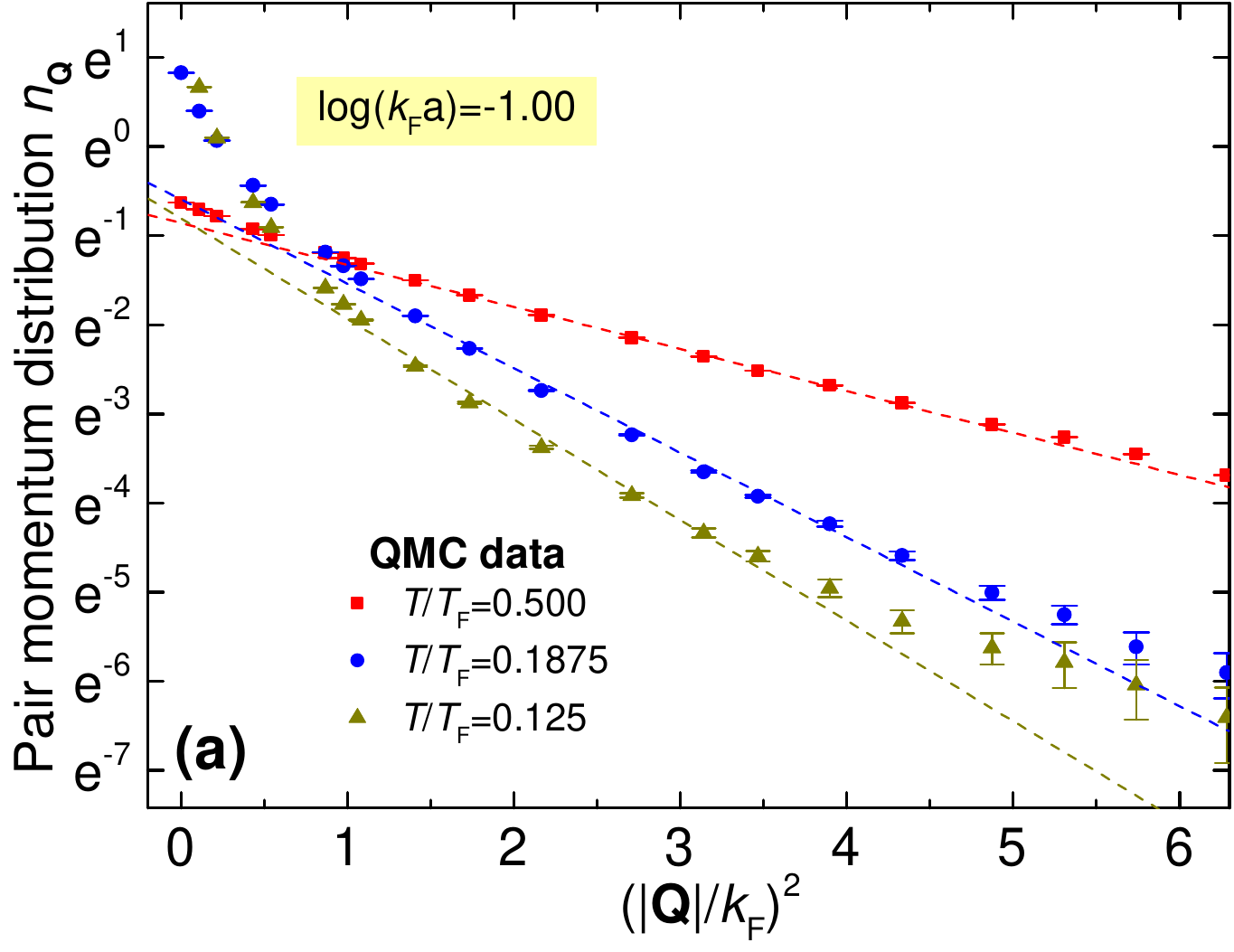} \hspace{0.3cm}
\includegraphics[width=0.46\columnwidth]{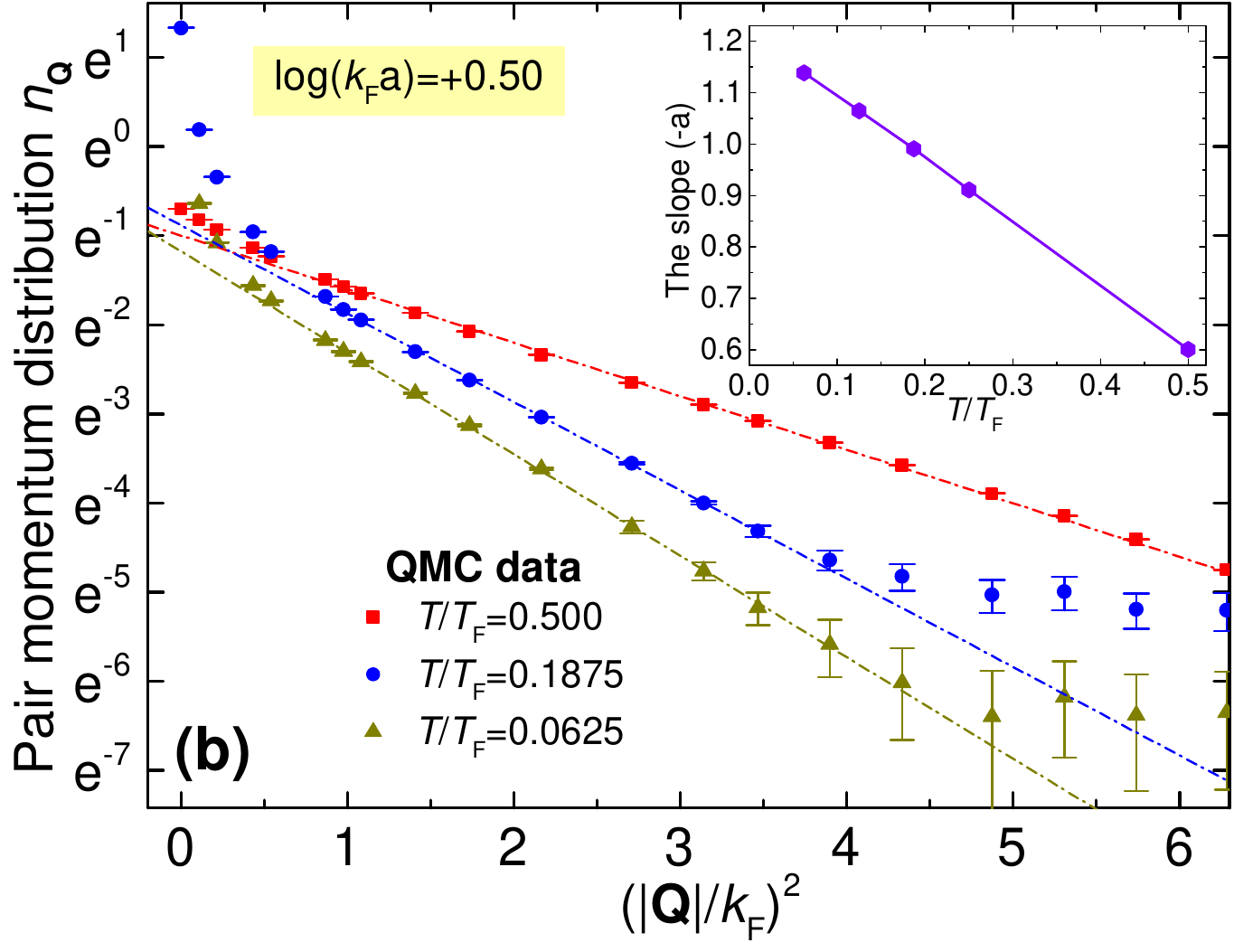}
\caption{\label{fig:PairnQ} The Cooper pair momentum distribution $n_{\mathbf{Q}}$ versus $(|\mathbf{Q}|/k_F)^2$ in semi-log plot for (a) $\log(k_Fa)=-1.00$ and (b) $\log(k_Fa)=+0.50$. The points are our QMC data, while the dashed lines are the corresponding $e^{ax+b}$ fitting [with $x=(|\mathbf{Q}|/k_F)^2$]. In panel (b), the slope $(-a)$ [in the semi-log plot] is plotted versus $T/T_F$. The system is $L=45,N_e=58$. }
\end{figure}

\section{III. Cooper pair momentum distribution}

\hspace{0.5cm} In the main text, we have shown the results of Cooper pair momentum distribution $n_{\mathbf{Q}}$ versus the center-of-mass momentum $\mathbf{Q}/k_F$. Here we present further discussion of these results in connection with the experimental results in Ref.~\onlinecite{Ries2015}. 

\hspace{0.5cm} In Fig.~\ref{fig:PairnQ}, we show the results of $n_{\mathbf{Q}}$ versus $(\mathbf{Q}/k_F)^2$ in the semi-log plots for both the BEC regime [$\log(k_Fa)=-1.00$] and the crossover regime [$\log(k_Fa)=+0.50$]. We see that, to within the resolution of the statistical errors, $\log(n_{\mathbf{Q}})$ shows a linear dependence on $(\mathbf{Q}/k_F)^2$, which is consistent with the experimental results~\cite{Ries2015}. The absolute value of the slope (which is $-a$ as indicated in Fig.~\ref{fig:PairnQ}) increases at lower temperatures. This is qualitatively consistent with the relation applied in Ref.~\onlinecite{Ries2015}, which was used to determine the temperature in the experiment in the BEC and crossover regimes.

\section{IV. The contact density $C/k_F^4$}

\hspace{0.5cm} In this section, we present the details of the formulas we applied to compute the contact density $C/k_F^4$ in the main text in both the AFQMC calculations and the mean-field theory. We also present AFQMC results of the contact extracted from the tail of the single-particle momentum distribution $n(\mathbf{k})$.

\subsection{A. The contact from double occupancy and from momentum distribution}

\hspace{0.5cm} For a general Fermi gas system, the contact can be computed from the Grand Potential $\Omega(\beta,V,\mu)$ as~\cite{Werner2012,Anderson2015,Jensen2020}
\begin{eqnarray}
\label{eq:CfromOmega00}
\mathcal{C} = 2\pi m \frac{\partial{\Omega(\beta,V,\mu)}}{\partial{(\ln a)}},
\end{eqnarray}
where $\Omega(\beta,V,\mu)$ is defined as $\Omega(\beta,V,\mu)=-\frac{1}{\beta}\ln Z(\beta,V,\mu)$ with $Z(\beta,V,\mu)=\text{Tr} e^{-\beta(\hat{H}+\mu\hat{N})}$ as the partition function of the system, and $a$ is the scattering length. So we have
\begin{eqnarray}
\label{eq:CfromOmega01}
\mathcal{C} = 2\pi m \frac{\partial{\Omega(\beta,V,\mu)}}{\partial{\ln a}}
= -\frac{2\pi m}{\beta}\frac{\partial{(\ln Z)}}{\partial{(\ln a)}}.
\end{eqnarray}
Writing the Hamiltonian of the system as $\hat{H}=\hat{K}+\hat{V}$ and $\hat{V}=U\hat{O}$, where $\hat{O}=\sum_{\mathbf{i}}n_{\mathbf{i}\uparrow}n_{\mathbf{i}\downarrow}$, we have
\begin{eqnarray}
\label{eq:PartialLnZ00}
\partial{(\ln Z)} = \frac{1}{Z}\partial{Z}
= \frac{1}{Z}\partial{[\text{Tr}e^{-\beta(\hat{H}+\mu\hat{N})}]}
= -\frac{\beta}{Z}\text{Tr}[e^{-\beta(\hat{H}+\mu\hat{N})}\partial{\hat{H}}]
= -\frac{\beta}{Z}\text{Tr}[e^{-\beta(\hat{H}+\mu\hat{N})}\partial{\hat{V}}],
\end{eqnarray}
and for $\partial{\hat{V}}$ we can obtain
\begin{eqnarray}
\label{eq:PartialV}
\partial{\hat{V}} = \partial{U}\hat{O} = (\partial{U}/U)\hat{V} = U\hat{V}(\partial{U}/U^2)
= -U\hat{V}\partial{(1/U)},
\end{eqnarray}
Thus we have 
\begin{eqnarray}
\label{eq:PartialLnZ01}
\partial{(\ln Z)} = \frac{1}{Z}\partial{Z}
= \beta U \partial{(1/U)} \cdot \frac{1}{Z} \text{Tr}[e^{-\beta(\hat{H}+\mu\hat{N})} \hat{V}]
= \beta U \langle\hat{V}\rangle \partial{(1/U)}
= \beta U^2 \langle\hat{O}\rangle \partial{(1/U)}.
\end{eqnarray}
Substituting the result of Eq.~(\ref{eq:PartialLnZ01}) into Eq.~(\ref{eq:CfromOmega01}), we have
\begin{eqnarray}
\label{eq:CfromOmega02}
\mathcal{C} = -\frac{2\pi m}{\beta}\frac{\partial{(\ln Z)}}{\partial{(\ln a)}}
= 2\pi m U^2 \langle\hat{O}\rangle \frac{\partial{(1/U)}}{\partial{(\ln a)}}.
\end{eqnarray} 
For the 2D system, we have \cite{Werner2012}
\begin{eqnarray}
\frac{\partial{(1/U)}}{\partial{(\ln a)}} = -\frac{m}{2\pi}.
\end{eqnarray}
Thus, we obtain the contact density 
\begin{eqnarray}
\label{eq:DouOccEq}
C = \frac{\mathcal{C}}{N_s} = \frac{m^2 U^2 \langle\hat{O}\rangle}{N_s}.
\end{eqnarray}
We also have $\langle\hat{O}\rangle=N_sD$ with $D$ denoting the double occupancy of the system. So we can further obtain $C/k_F^4$ as
\begin{eqnarray}
\label{eq:Result0}
\frac{C}{k_F^4} = \frac{m^2 U^2 D}{4\pi^2n^2}.
\end{eqnarray}

\hspace{0.5cm} We can also compute the contact from the asymptotic behavior~\cite{Werner2012} as 
\begin{eqnarray}
\label{eq:NofkAndContact}
\lim_{k/k_F\to+\infty} n(\mathbf{k})=\frac{C}{k^4}=\frac{C}{k_F^4}\frac{1}{(k/k_F)^4},
\end{eqnarray}

\subsection{B. The contact from BCS mean-field theory}

\hspace{0.5cm} The BCS mean-field Hamiltonian for the attractive Hubbard model can be written as
\begin{eqnarray}
\hat{H}_{\text{BCS}} 
&=& \sum_{\mathbf{k}}\Big[ (\xi_{\mathbf{k}\uparrow}c_{\mathbf{k}\uparrow}^+c_{\mathbf{k}\uparrow} - \xi_{\mathbf{k}\downarrow}c_{-\mathbf{k}\downarrow}c_{-\mathbf{k}\downarrow}^+ ) - U\Delta (c_{\mathbf{k}\uparrow}^+c_{\mathbf{-k}\downarrow}^+ + c_{\mathbf{-k}\downarrow}c_{\mathbf{k}\uparrow}) \Big] + \sum_{\mathbf{k}}\xi_{\mathbf{k}\downarrow} \\ \nonumber
&=& \sum_{\mathbf{k}}[ E_+(\mathbf{k})\alpha_{\mathbf{k}}^+\alpha_{\mathbf{k}} - E_-(\mathbf{k})\beta_{\mathbf{k}}^+\beta_{\mathbf{k}} ] + \sum_{\mathbf{k}}\xi_{\mathbf{k}\downarrow},
\end{eqnarray}
where $\Delta=\langle c_{\mathbf{i}\uparrow}^+c_{\mathbf{i}\downarrow}^+\rangle = \langle c_{\mathbf{i}\downarrow}c_{\mathbf{i}\uparrow}\rangle$ as the mean-field order parameter, and $E_{\pm}(\mathbf{k})=[(\xi_{\mathbf{k}\uparrow}-\xi_{\mathbf{k}\downarrow})\pm\sqrt{(\xi_{\mathbf{k}\uparrow}+\xi_{\mathbf{k}\downarrow})^2+4(U\Delta)^2}]/2$ with $\xi_{\mathbf{k}\sigma}=\varepsilon_{\mathbf{k}\sigma}+\mu$. For our case, we have $\xi_{\mathbf{k}\uparrow}=\xi_{\mathbf{k}\downarrow}=\xi_{\mathbf{k}}$. The solution of the BCS mean-field theory gives the following self-consistent equations for $\Delta$ and $N_e$ 
\begin{eqnarray}
N_s - N_e = \sum_{\mathbf{k}}\frac{\xi_{\mathbf{k}}}{\sqrt{\xi_{\mathbf{k}}^2+(U\Delta)^2}}\Big(1 - \frac{2}{1+e^{\beta E_{+}(\mathbf{k})}} \Big)
\hspace{1.5cm}
\Delta = \frac{1}{2N_s}\sum_{\mathbf{k}}\frac{U\Delta}{\sqrt{\xi_{\mathbf{k}}^2+(U\Delta)^2}}\Big(1 - \frac{2}{1+e^{\beta E_{+}(\mathbf{k})}} \Big).
\end{eqnarray}
Thus, we can rewrite the BCS mean-field Hamiltonian as $\hat{H}_{\text{BCS}}=\hat{K}+\hat{V}$ and $\hat{V}=U\hat{O}$ with $\hat{O}=\sum_{\mathbf{i}}(c_{\mathbf{i}\uparrow}^+c_{\mathbf{i}\downarrow}^+\langle c_{\mathbf{i}\downarrow}c_{\mathbf{i}\uparrow}\rangle + \langle c_{\mathbf{i}\uparrow}^+c_{\mathbf{i}\downarrow}^+\rangle c_{\mathbf{i}\downarrow}c_{\mathbf{i}\uparrow} - \langle c_{\mathbf{i}\uparrow}^+c_{\mathbf{i}\downarrow}^+\rangle\langle c_{\mathbf{i}\downarrow}c_{\mathbf{i}\uparrow}\rangle)$ and $\langle\hat{O}\rangle=N_s\Delta^2$. The contact density from BCS can be readily computed:
\begin{eqnarray}
\label{eq:Result1}
\frac{C_{\text{BCS}}}{k_F^4} = \frac{m^2 U^2 \Delta^2}{4\pi^2 n^2}.
\end{eqnarray}

\hspace{0.5cm} The single-particle momentum distribution $n(\mathbf{k})$ from BCS is given by
\begin{eqnarray}
\label{eq:BCSNofk}
n(\mathbf{k}) = \frac{1}{2}\Big[ 1 - \frac{k^2/2m+\mu}{\sqrt{(k^2/2m+\mu)^2+(U\Delta)^2}} \Big(1 - \frac{2}{e^{\beta E(\mathbf{k})}+1} \Big)  \Big].
\end{eqnarray}
Here, we have applied the quadratic dispersion of $\varepsilon_{\mathbf{k}}=k^2/2m$ (we set $\hbar=1$) with $k=|\mathbf{k}|$, and the BCS energy dispersion relation $E(\mathbf{k})=E_{+}(\mathbf{k})=\sqrt{(k^2/2m+\mu)^2+(U\Delta)^2}$. 

\hspace{0.5cm} At $T=0$,  the BCS result simplifies to
\begin{eqnarray}
\label{eq:BCSNofkZeroT}
n(\mathbf{k}) = \frac{1}{2}\Big[ 1 - \frac{k^2/2m+\mu}{\sqrt{(k^2/2m+\mu)^2+(U\Delta)^2}} \Big].
\end{eqnarray}
Expanding the second term in terms of $1/k$ for the asymptotic behavior in $k$, we obtain
\begin{eqnarray}
\label{eq:Taylor00}
\frac{k^2/2m+\mu}{\sqrt{(k^2/2m+\mu)^2+(U\Delta)^2}}
&=& 1 - \frac{(U\Delta)^2}{2}\Big(\frac{2m}{k^2}\Big)^2\Big(1+\frac{2m\mu}{k^2}\Big)^{-2} + \frac{3(U\Delta)^4}{8}\Big(\frac{2m}{k^2}\Big)^4\Big(1+\frac{2m\mu}{k^2}\Big)^{-4} + \cdots\cdots.
\end{eqnarray}
This leads to the asymptotic behavior of $n(\mathbf{k})$ as
\begin{eqnarray}
\label{eq:NkZeroTkInf00}
n(\mathbf{k}) 
\stackrel{k\to\infty}{\Longrightarrow} \frac{m^2U^2\Delta^2}{k^4} + \mathcal{O}(k^{-6})
= \frac{m^2U^2\Delta^2}{4\pi^2n^2}\frac{1}{(k/k_F)^4} + \mathcal{O}[(k/k_F)^{-6}]\,.
\end{eqnarray}
Comparing to Eq.~(\ref{eq:NofkAndContact}), we have the contact density in BCS theory  at $T=0$:
\begin{eqnarray}
\label{eq:BCSContact00}
\frac{C}{k_F^4} = \frac{m^2U^2\Delta^2}{4\pi^2n^2}.
\end{eqnarray}
We next consider the finite-temperatire case $T>0$ in BCS theory. Again, applying the Taylor expansion, we have 
\begin{eqnarray}
\label{eq:FiniteTFactor}
1 - \frac{2}{e^{\beta E(\mathbf{k})}+1} 
&=& 1 - 2e^{-\beta E(\mathbf{k})} + \mathcal{O}(e^{-2\beta E(\mathbf{k})})  \\ \nonumber
&=& 1 - 2\exp\Big(-\beta\frac{k^2}{2m}\sqrt{\Big(1+\frac{2m\mu}{k^2}\Big)^2 + \Big(\frac{2mU\Delta}{k^2}}\Big)^2  \Big) + \mathcal{O}(e^{-2\beta E(\mathbf{k})}),
\end{eqnarray}
where we see that the leading order of the second term is $e^{-\beta k^2}$, which decays faster than the power-law decay of $1/k^2$ for sufficiently low temperatures. Thus, the second term of the finite-temperature factor in Eq.~(\ref{eq:FiniteTFactor}) will not change the leading behavior of $1/k^4$ in $n(\mathbf{k})$ at $T=0$ given in Eq.~(\ref{eq:NkZeroTkInf00}). As a result, the contact density formula in Eq.~(\ref{eq:BCSContact00}) remains valid for finite temperatures and the temperature effect is already incorporated into the BCS mean-field order parameter $\Delta$, which can change with $T$. 

\hspace{0.5cm} We can see that these two methods for computing $C/k_F^4$ result in exactly the same formulas in Eq.~(\ref{eq:Result1}) and Eq.~(\ref{eq:BCSContact00}) for BCS mean-field theory. 

\subsection{C. Some AFQMC results}

\hspace{0.5cm} In the AFQMC simulations, we have calculated the contact density $C/k_F^4$ using both methods: from double occupancy as in Eq.~(\ref{eq:Result0}), and from the asymptotic behavior of $n(\mathbf{k})$ as in Eq.~(\ref{eq:NofkAndContact}), to make a self-contained check.

\begin{figure}[ht]
\centering
\includegraphics[width=0.45\columnwidth]{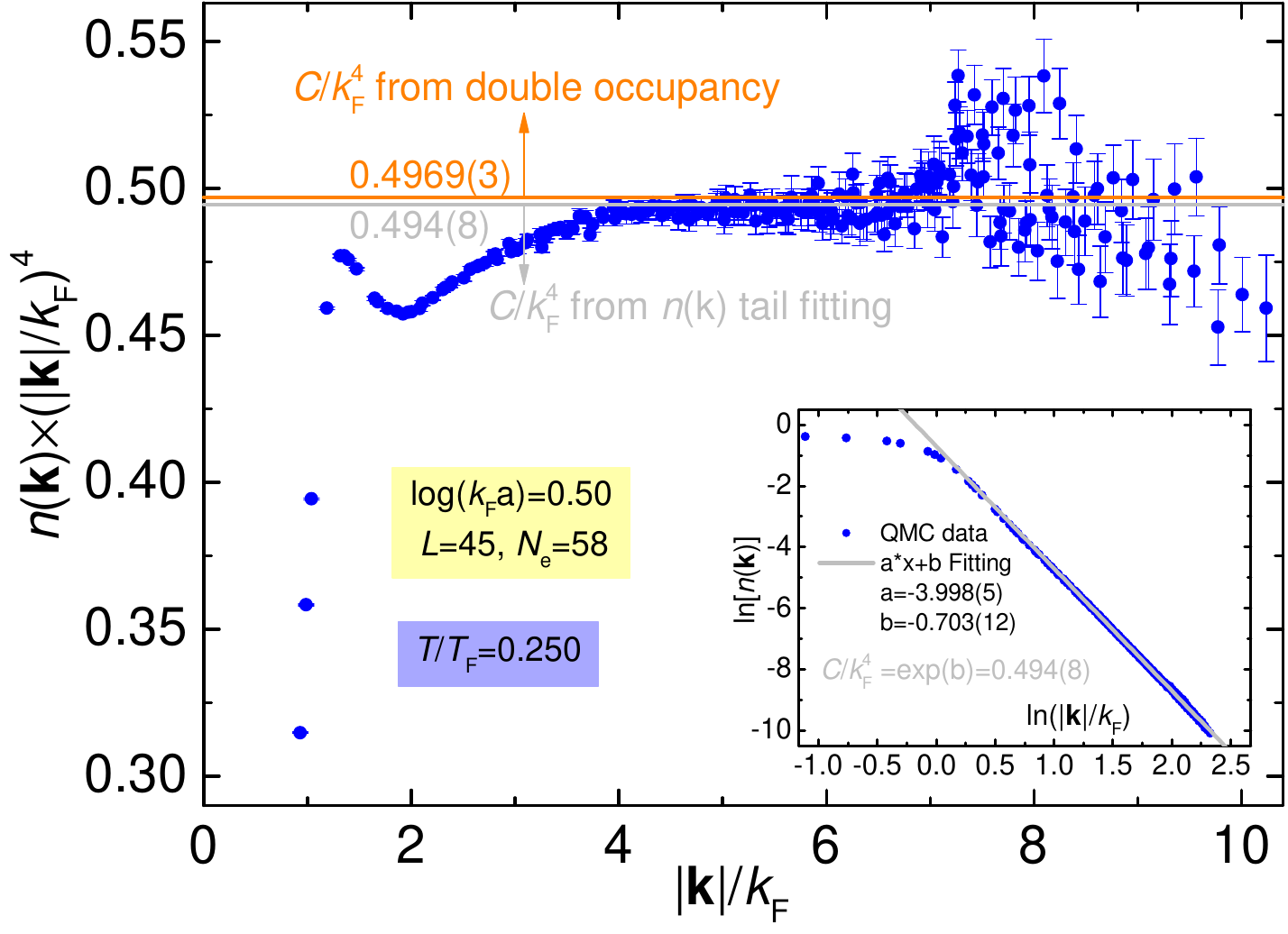} 
\hspace{0.2cm}
\includegraphics[width=0.45\columnwidth]{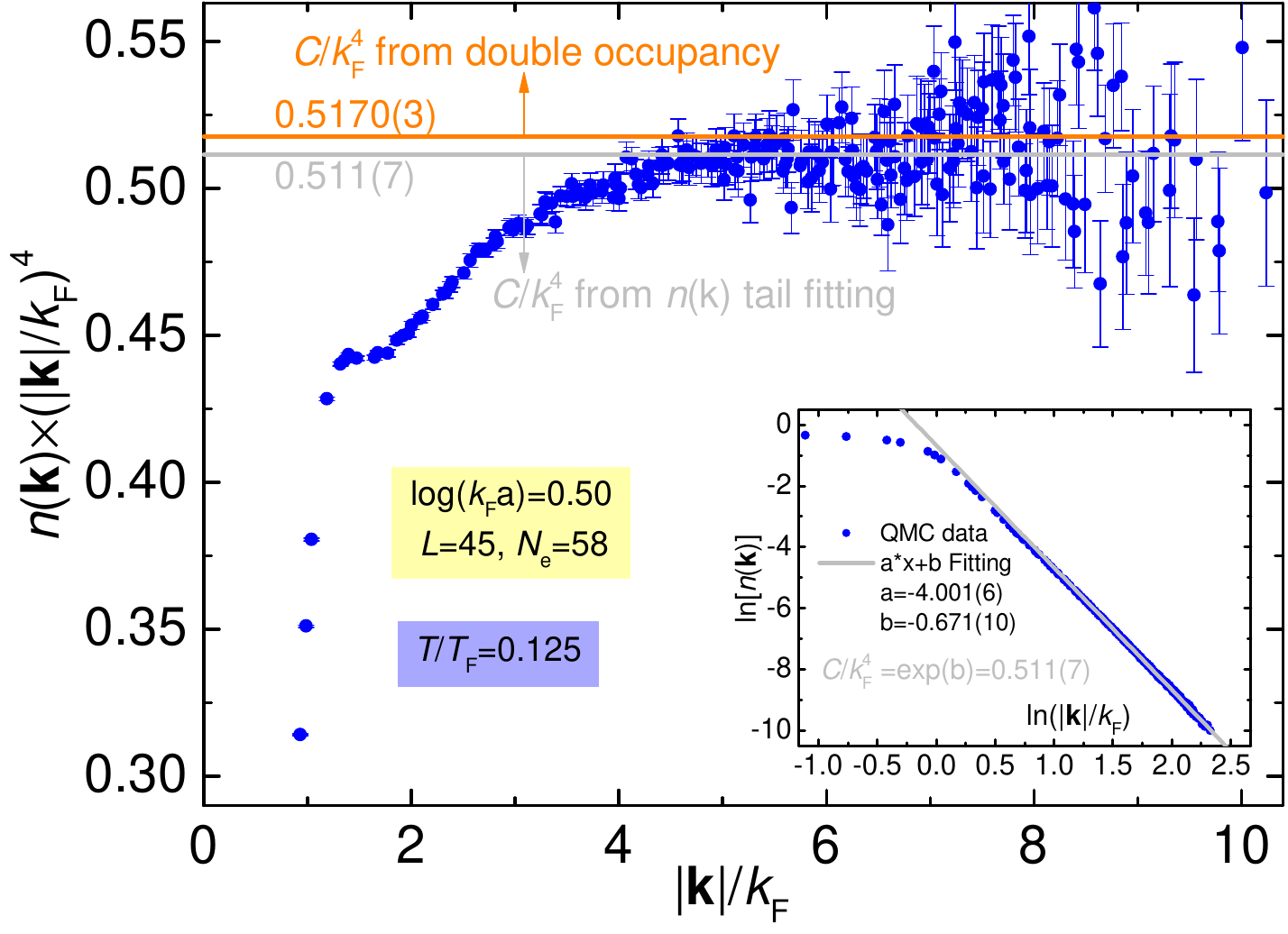} \\
\includegraphics[width=0.45\columnwidth]{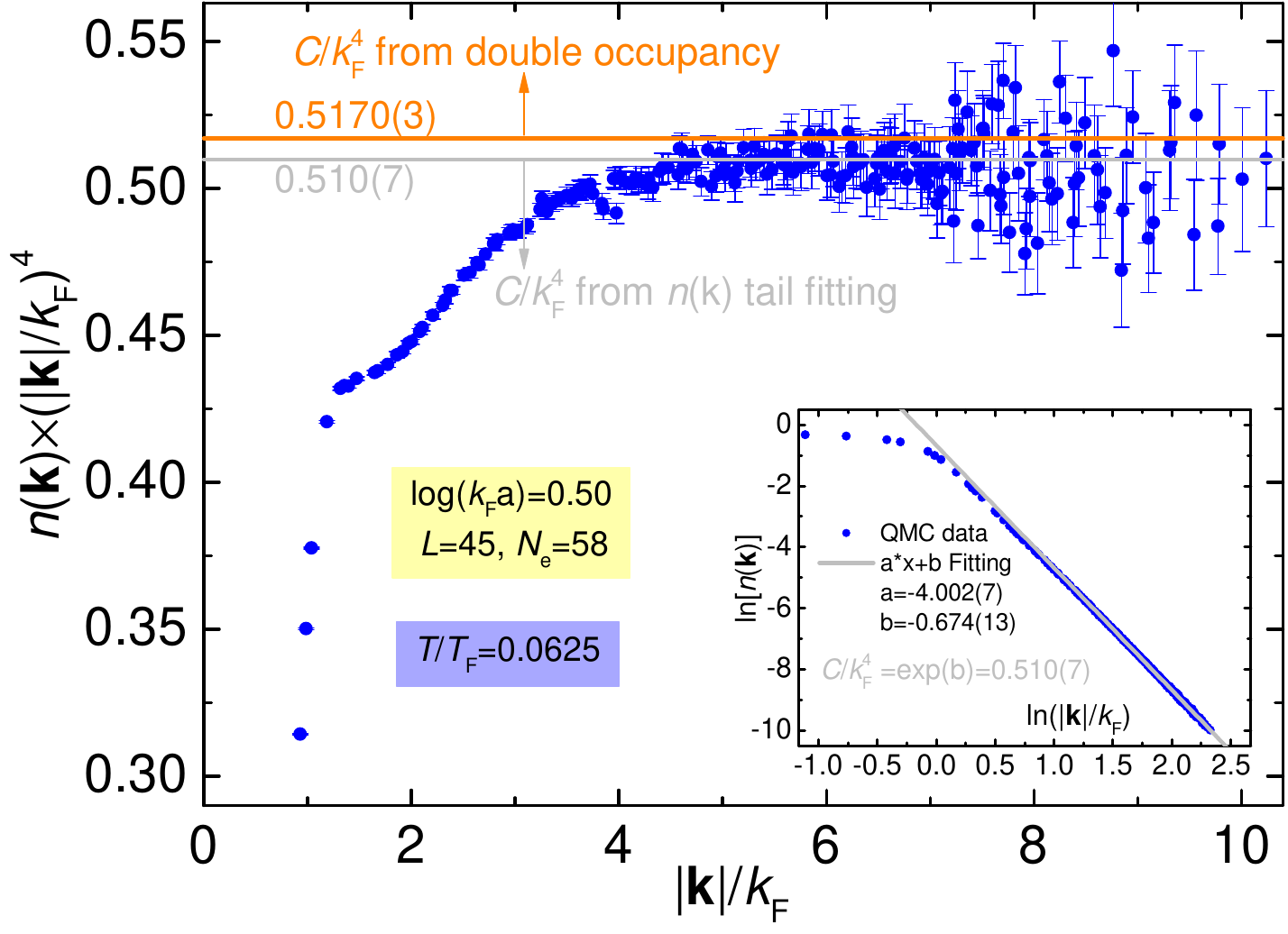}
\hspace{0.2cm}
\includegraphics[width=0.45\columnwidth]{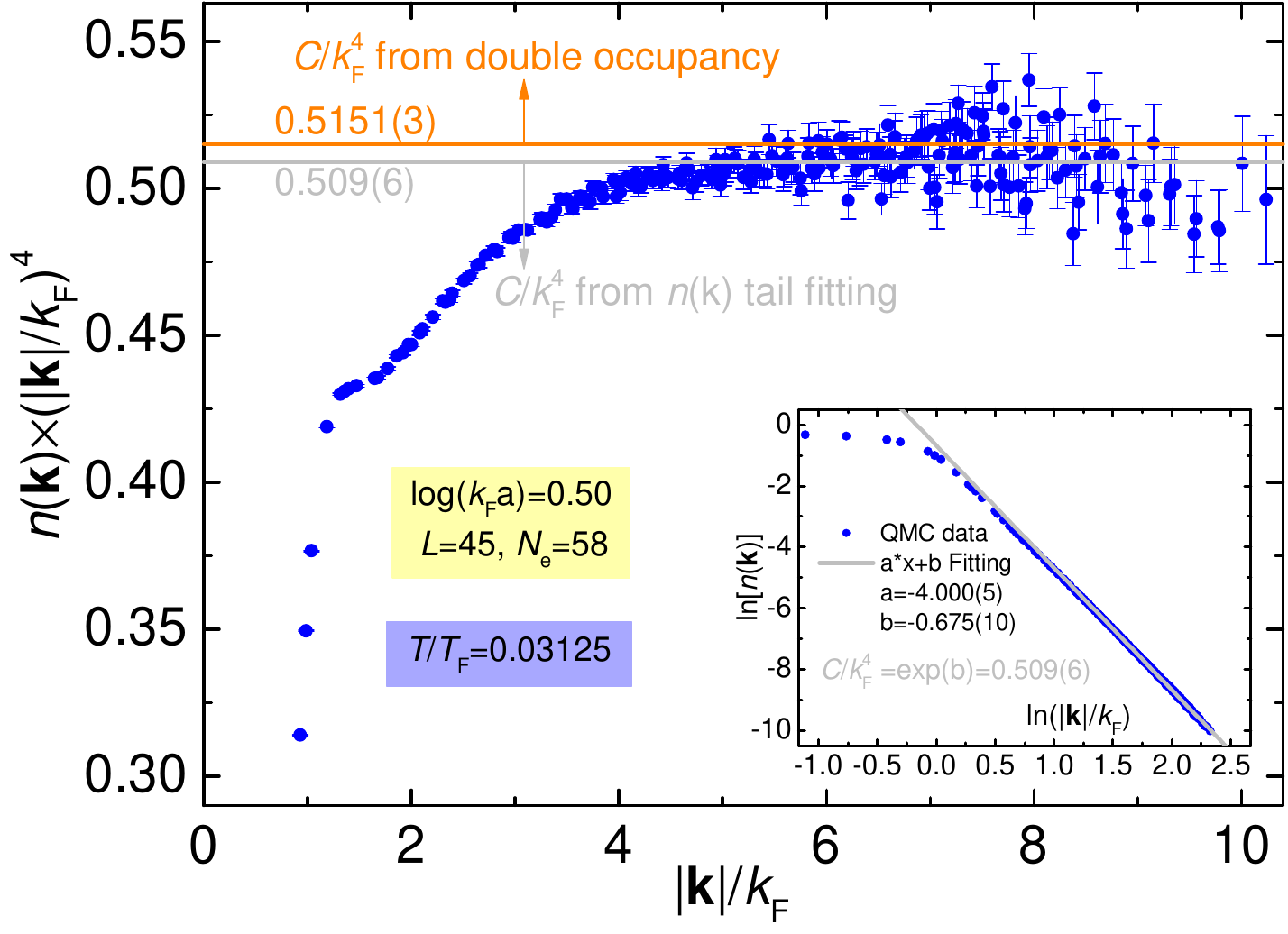}
\caption{\label{fig:ContactDensity} Extracting the contact density $C/k_F^4$ from the tail of fermion momentum distribution function $n(\mathbf{k})$, for $\log(k_Fa)=+0.50$ with temperatures $T/T_F=0.25, 0.125, 0.0625, 0.0325$. The main plots show the results of $n(\mathbf{k})\times(|\mathbf{k}|/k_F)^4$, in which pleateus exist at high momenta. In the insets, the fittings for the tail of $n(\mathbf{k})$ using $a/(|\mathbf{k}|/k_F)^4+b$ are demonstrated in the log-log plot. The system is $L=45,N_e=58$ with quadratic dispersion. }
\end{figure}

In Fig.~\ref{fig:ContactDensity}, we show the comparisons of $C/k_F^4$ extracted from Eq.~(\ref{eq:Result0}) and Eq.~(\ref{eq:NofkAndContact}) for $\log(k_Fa)=+0.50$ with four characteristic temperatures. In the main plots, the pleateus of $n(\mathbf{k})\times(|\mathbf{k}|/k_F)^4$ at high momenta exist as expected. The insets shows the linear fittings of $\log[n(\mathbf{k})]$ versus $\log(|\mathbf{k}|/k_F)$ at large $|\mathbf{k}|/k_F$ region with slopes very close to the theoretical value of $4$, with high quality. The resulting statistical error bars are different, with the contact from the momentum distribution considerably more noisy than from double-occupancy. However the contact density $C/k_F^4$ extracted from the two different methods agree very well.

\section{V. AFQMC simulation setup and data for reproducibility}

\hspace{0.5cm} In this section, we present the details of the AFQMC simulation setup in this work, and show data of condensate fraction and double occupancy in a number of systems to allow the reproducibility. 

\hspace{0.5cm} Evolving from the BCS side to the BEC side, the interaction strength, namely the value of $U$, can change from $-1.0$ to $-10.0$ in practical simulations. Thus, we decrease the $\Delta\tau$ from $0.04$ to $0.01$ crosspondingly to keep the Trotter error negligible. Applying the speedup algorithm~\cite{YuanYao2019}, we typically apply the cutoff $\epsilon=10^{-5}$ to aviod truncation error. The statistics also varies with the change of interaction strength, but the relative error of typical quantities, such as condensate fraction and double occupancy, are generally kept less than $1\%$.

\hspace{0.5cm} The condensate fraction is the core quantity we applied to determine the BKT transition temperatures, and the double occupancy is used to compute the contact (as shown in Eq.~(\ref{eq:Result0})) in this work. So in the following, we present detailed data for these two quantities to allow for data reproducibility. 

\begin{table}[h!]
\caption{\label{tab:table1} The results of condensate fraction and double occupancy for different systems $L=15,19,23$ with $\log(k_Fa)=+2.346573590279973$ and $N_e=58$ at the temperature $T/T_F=0.0625$. The values of $U$ and $\beta$ are also presented. The Trotter discritization $\Delta\tau=0.04$ is used. }
\begin{tabular}{|p{2.6cm}<{\centering}|p{2.5cm}<{\centering}|p{2.5cm}<{\centering}|p{3.3cm}<{\centering}|p{2.8cm}<{\centering}|}
\hline
\centering
Linear system size &  $U$  &  $\beta$   &   condensate Fraction  &  Double occupancy \\
\hline
$L=15$ & $-3.37587752648$ & $9.87858267467$  & 0.0537(1) & 0.041740(4) \\
\hline
$L=19$ & $-3.17429584008$ & $15.84963709136$ & 0.0547(2) & 0.018476(4) \\
\hline
$L=23$ & $-3.02815476725$ & $23.22564548844$ & 0.0539(5) & 0.009463(2) \\
\hline
\end{tabular}
\end{table}

\begin{table}[h!]
\caption{\label{tab:table2} The results of condensate fraction and double occupancy for $L=21,25,27$ with $\log(k_Fa)=+0.346573590279973$ and $N_e=58$ at the temperature $T/T_F=0.25$. The values of $U$ and $\beta$ are also presented. The Trotter discritization $\Delta\tau=0.02$ is used. }
\begin{tabular}{|p{2.6cm}<{\centering}|p{2.5cm}<{\centering}|p{2.5cm}<{\centering}|p{3.3cm}<{\centering}|p{2.8cm}<{\centering}|}
\hline
\centering
Linear system size &  $U$  &  $\beta$   &   condensate Fraction  &  Double occupancy \\
\hline
$L=21$ & $-7.94957849650$ & $4.84050551059$ & 0.07849(7) & 0.031303(9) \\
\hline
$L=25$ & $-7.15986516344$ & $6.86012685741$ & 0.08527(5) & 0.018687(4) \\
\hline
$L=27$ & $-6.85909634017$ & $8.00165196648$ & 0.08744(7) & 0.014817(4) \\
\hline
\end{tabular}
\end{table}

Table~\ref{tab:table1} lists the results of $\log(k_Fa)=+2.346573590279973$ and $N_e=58$ at the temperature $T/T_F=0.0625$, for linear system size $L=15\sim 25$, with applied $\Delta\tau=0.04$. In total 500 sweeps are used to equilibrate the Markov chain, after which around $10^4$ sweeps are used to perform the measurements. The numerical stablization is applied every 18 imaginary-time slices with cutoff $\epsilon=10^{-5}$ for the low-rank factorization~\cite{YuanYao2019}. Since the interaction strength is sufficiently small, we apply the Hubbard-Stratonovich transformation into density channel for this case. 

Table~\ref{tab:table2} lists the results of $\log(k_Fa)=+0.346573590279973$ and $N_e=58$ at the temperature $T/T_F=0.25$, for linear system size $L=21\sim 27$, with applied $\Delta\tau=0.02$. 200 sweeps are used to equilibrate the Markov chain with around $4\times 10^3$ sweeps used to perform the measurements afterwards. Numerical stablization is applied every 18 imaginary-time slices with cutoff $\epsilon=10^{-5}$. Since the interaction strength is quite large, we apply the Hubbard-Stratonovich transformation into spin-$\hat{s}_z$ channel which reduces the autocorrelation time. 

\end{document}